\pdfoutput=1
\documentclass[manuscript,screen,nonacm]{acmart}

\AtBeginDocument{%
  }

\makeatletter

\let\digamma\relax

\makeatother

\usepackage{amssymb}
\usepackage{algorithm}
\usepackage{algpseudocode}
\usepackage{alltt}
\usepackage{amsmath}
\usepackage{booktabs}
\usepackage{caption}
\usepackage[capitalise, nameinlink, noabbrev]{cleveref}
\usepackage{color}
\usepackage{colortbl}
\usepackage[inline]{enumitem}
\usepackage{fancybox}
\usepackage{fp}
\usepackage{graphicx}
\usepackage{hyperref}
\usepackage{listings}
\usepackage{multirow}
\usepackage{pifont}
\usepackage{subcaption}
\usepackage{textcomp}
\usepackage{url}
\usepackage{wrapfig}
\usepackage{xcolor}
\usepackage{xspace}
\usepackage{todonotes}
\usepackage{tikz}
\usepackage[most]{tcolorbox}
\usetikzlibrary{arrows.meta,positioning,calc,fit}

\def\DEFECTS4J{{\scshape{}Defects4J}\xspace}
\def\EFDD{{\scshape{}EFDD}\xspace}

\def\SFLKIT{{\scshape{}SFLKit}\xspace}

\def\TESTS4PY{{\scshape{}Tests4Py}\xspace}

\def\WORD2VEC{{\smaller{}WORD2VEC}\xspace}

\def\FORECAST{{\smaller{}FORECAST}\xspace}
\def\SWELITE{{\scshape{}SWE-bench Lite}\xspace}
\def\AUTOCODEROVER{{\scshape{}AutoCodeRover}\xspace}
\def\AGENTLESS{{\scshape{}Agentless}\xspace}
\def\SWEAGENT{{\scshape{}SWE-agent}\xspace}
\def\REPAIRAGENT{{\scshape{}RepairAgent}\xspace}
\def\MINISWEAGENT{{\scshape{}mini-SWE-agent}\xspace}
\def\HAIKU{{\scshape{}Claude Haiku~4.5}\xspace}
\def\HAIKUSHORT{{\scshape{}Haiku~4.5}\xspace}
\def\QWEN{{\scshape{}Qwen3.5}\xspace}

\def\SWEBENCH{{\scshape{}SWE-bench}\xspace}
\def\ANTHROPIC{{\scshape{}Anthropic}\xspace}

\newcommand{\nInstances}{300}
\newcommand{\nProjects}{12}
\newcommand{\nAttempts}{6{,}000}
\newcommand{\nFailingTraced}{1{,}040}
\newcommand{\nPassingTraced}{16{,}720}
\newcommand{\nPerfectChains}{274}
\newcommand{\nFixedFileChains}{166}
\newcommand{\diagMedianMinutes}{40}
\newcommand{\haikuBase}{48.5\%}
\newcommand{\haikuForecast}{70.9\%}
\newcommand{\qwenBase}{31.2\%}
\newcommand{\qwenForecast}{56.1\%}
\newcommand{\haikuGain}{22.4}
\newcommand{\qwenGain}{24.9}
\newcommand{\haikuGainRel}{46\%}
\newcommand{\qwenGainRel}{80\%}
\newcommand{\haikuGained}{371}
\newcommand{\haikuLost}{35}
\newcommand{\qwenGained}{387}
\newcommand{\qwenLost}{13}

\newcommand{\haikuGainsPerLoss}{11}
\newcommand{\qwenGainsPerLoss}{30}
\newcommand{\raGainHaiku}{45.0}
\newcommand{\raGainQwen}{40.0}
\newcommand{\minAgentGain}{8.3}
\newcommand{\maxAgentGain}{\raGainHaiku}
\newcommand{\haikuTokenDrop}{60\%}
\newcommand{\qwenTokenDrop}{50\%}
\newcommand{\haikuCostDrop}{42\%}

\definecolor{rltred}{rgb}{0.5,0,0}
\definecolor{rltgreen}{rgb}{0,0.5,0}
\definecolor{rltblue}{rgb}{0,0,0.5}
\hypersetup{
colorlinks=true,
urlcolor=rltblue,
linkcolor=rltred,
citecolor=rltgreen,
}

\newcommand{\mathid}[1]{\textit{#1}}
\newcommand{\codeid}[1]{\texttt{#1}}
\def\|#1|{\mathid{#1}}
\def\<#1>{\codeid{#1}}

\definecolor{deepblue}{rgb}{0,0,0.5}
\definecolor{deepred}{rgb}{0.6,0,0}
\definecolor{deepgreen}{rgb}{0,0.5,0}
\definecolor{deeporange}{rgb}{0.8,0.45,0}
\definecolor{Grey}{rgb}{0.5,0.5,0.5}
\definecolor{LightGrey}{rgb}{0.9,0.9,0.9}
\definecolor{forecastblue}{HTML}{1A85FF}
\colorlet{tablerow}{forecastblue!8}
\definecolor{Green}{rgb}{0.0,0.6,0.0}
\definecolor{Red}{rgb}{0.6,0.0,0.0}
\definecolor{Blue}{rgb}{0.0,0.0,0.6}

\definecolor{DarkBlue}{rgb}{0.0859, 0.308, 0.523}
\definecolor{DarkOrange}{rgb}{0.8, 0.4, 0.0}
\definecolor{DarkGreen}{rgb}{0.00,0.40,0.00}
\definecolor{ScarletRed}{rgb}{0.60,0.00,0.00}
\definecolor{AlmostWhite}{rgb}{0.80,0.80,0.80}
\definecolor{Gray}{gray}{0.85}

\definecolor{row}{rgb}{0.8627, 0.9098, 0.9804}

\definecolor{Cornsilk}{rgb}{0.98, 0.94, 0.9}

\newcommand{\PASS}{\text{\color{Green}\ding{52}}\xspace}
\newcommand{\FAIL}{\text{\color{Red}\ding{56}}\xspace}

\newcolumntype{a}{>{\columncolor{TableHeader}}l}

\newcounter{confusions}
\let\backslashpipe=\|
\let\backslashstar=\*  %
\let\backslashangle=\<
\def\*#1*{\codeid{#1}}  %

\newcommand{\maketodo}[2]{\expandafter\newcommand\csname #1\endcsname[1]{\todo[bordercolor=#2!80!black,color=#2]{\textbf{\MakeUppercase #1:} ##1}\xspace}}

\presetkeys{todonotes}{fancyline}{}
\definecolor{todored}{rgb}{1, 0.6, 0.6}
\definecolor{todoorange}{rgb}{1, 0.8, 0.4}
\definecolor{todoblue}{rgb}{0.4, 0.8, 1.0}
\definecolor{todogreen}{rgb}{0.8, 1.0, 0.4}
\definecolor{todopurple}{RGB}{255, 100, 127}

\maketodo{NOTE}{todogreen}
\maketodo{TODO}{todored}
\maketodo{CHECK}{todoblue}
\newcommand{\DONE}[1]{} %
\newcommand{\WONTFIX}[1]{} %
\newcommand{\ALSO}[1]{} %

\DeclareRobustCommand{\circnum}[1]{\tikz[baseline=(c.base)]\node[circle, draw, inner sep=0pt, minimum size=3.5mm, font=\tiny] (c) {#1};}

\newtcolorbox{finding}[1][Finding]{%
  breakable, enhanced, arc=1.5pt, colback=black!4, colframe=black!60,
  boxrule=0.5pt, left=5pt, right=5pt, top=2.5pt, bottom=2.5pt,
  fonttitle=\bfseries\footnotesize, coltitle=white, title={#1},
  before skip=5pt, after skip=5pt, width=\linewidth}

\newlist{questions}{enumerate}{1}
\setlist[questions,1]{label=\bfseries RQ\arabic*:,ref=RQ\arabic*,leftmargin=3\parindent}
\crefname{question}{}{}
\Crefname{question}{}{}

\title{Improving Repair Agents with Cause-Chain Diagnoses}

\author{Marius Smytzek}
\orcid{0000-0002-4899-9031}
\affiliation{%
 \institution{CISPA Helmholtz Center for Information Security}
 \city{Saarbr{\"u}cken}
 \country{Germany}
}
\email{marius.smytzek@cispa.de}

\author{Martin Eberlein}
\orcid{0000-0003-4268-7632}
\affiliation{%
 \institution{Humboldt-Universit{\"a}t zu Berlin}
 \city{Berlin}
 \country{Germany}
}
\email{martin.eberlein@hu-berlin.de}

\author{Lars Grunske}
\orcid{0000-0002-8747-3745}
\affiliation{%
 \institution{Humboldt-Universit{\"a}t zu Berlin}
 \city{Berlin}
 \country{Germany}
}
\email{grunske@hu-berlin.de}

\author{Andreas Zeller}
\orcid{0000-0003-4719-8803}
\affiliation{%
 \institution{CISPA Helmholtz Center for Information Security}
 \city{Saarbr{\"u}cken}
 \country{Germany}
}
\email{zeller@cispa.de}



\begin{abstract}
    Repair agents resolve a growing share of real-world issues.
    Starting with the same information as a developer reading a bug report, they have to find out how the program fails before actually fixing it.

    We present \FORECAST{}, which provides repair agents with a diagnosis of the failure.
    From the executions of passing and failing tests, \FORECAST{} infers a \emph{cause chain}, a short sequence of execution features, such as conditions or data flow, that explains how the failing runs diverge.
    \FORECAST{} appends this cause chain to the issue an agent receives.
    It thus works with any agent that builds its prompts from the issue, without modifying the agent.

    We evaluate \FORECAST{} on all $\nInstances$ issues of \SWELITE{} with five repair agents and two models, the commercial \HAIKU{} and the open-weight \QWEN{}.
    With the cause-chain diagnosis, every agent resolves significantly more issues.
    Over all agents, the share of resolved issues rises from $\haikuBase$ to $\haikuForecast$ for \HAIKUSHORT{} and from $\qwenBase$ to $\qwenForecast$ for \QWEN{}.
    The open-weight model with the diagnosis thus surpasses the commercial model without it.
    The gains persist across repeated runs and for agents that already run the tests the diagnosis is derived from.
\end{abstract}

\begin{document}

\maketitle

\section{Introduction}%
\label{sec:introduction}

\begin{wrapfigure}{r}{0.5\textwidth}
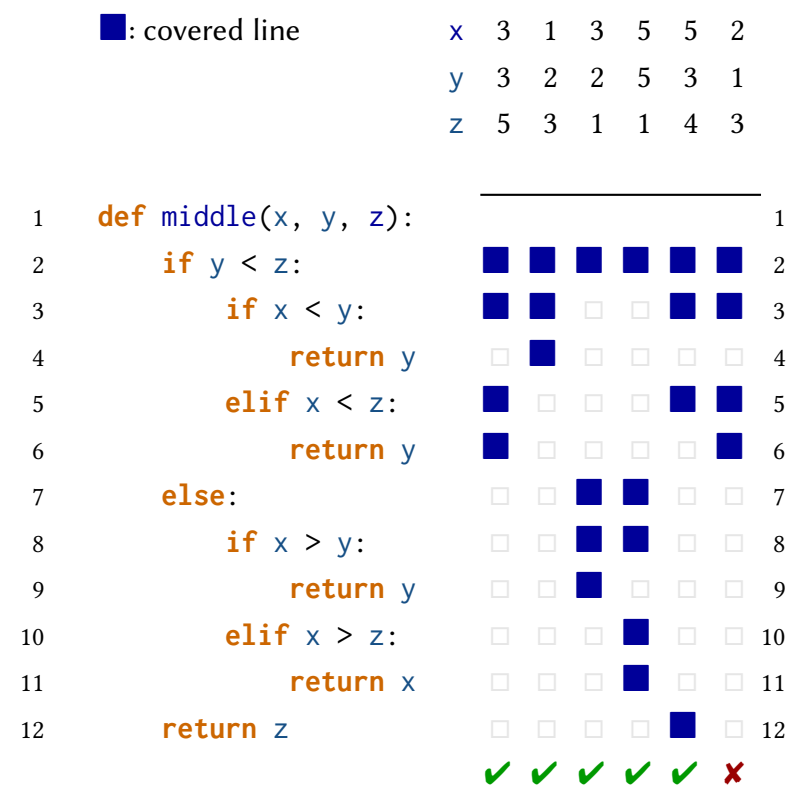

    \def\*{{\color{Blue}$\blacksquare$}}
    \def\+{{\color{Blue}$\blacksquare$}}
    \def\-{{\color{LightGrey}$\Box$}}
    \def\ind{\qquad}
    \scriptsize
    \centering
    \begin{tabular}{@{}>{\tiny}r>{\tt}l@{\quad}l@{\ \ }r@{\ \ }r@{\ \ }r@{\ \  }r@{\ \ }r@{\ \ }r@{\ \ }>{\tiny}r}
    \\
    & \textsf{\color{black}\*: covered line}
         & \color{Blue}\texttt{x}      & 3   & 1   & 3   & 5   & 5   & 2  \\
       & & \color{DarkBlue}\texttt{y}  & 3   & 2   & 2   & 5   & 3   & 1  \\
       & &  \color{DarkBlue}\texttt{z} & 5   & 3   & 1   & 1   & 4   & 3  \\
       \\ \cline{4-9}
    1  & \textbf{\color{DarkOrange}def} {\color{Blue}middle}({\color{DarkBlue}x}, {\color{DarkBlue}y}, {\color{Blue}z}):  & & & & & & & & 1 \\
    2  & \ind  \textbf{\color{DarkOrange}if} {\color{DarkBlue}y} < {\color{DarkBlue}z}:
    & & \*  & \*  & \*  & \*  & \*  & \*  & 2 \\ %
    3  & \ind \ind  \textbf{\color{DarkOrange}if} {\color{DarkBlue}x} < {\color{DarkBlue}y}:
    & & \*  & \*  & \-  & \-  & \*  & \*  & 3 \\ %
    4  & \ind \ind \ind  \textbf{\color{DarkOrange}return} {\color{DarkBlue}y}
    & & \-  & \*  & \-  & \-  & \-  & \-  & 4 \\ %
    5  & \ind \ind  \textbf{\color{DarkOrange}elif} {\color{DarkBlue}x} < {\color{DarkBlue}z}:
    & & \*  & \-  & \-  & \-  & \*  & \*  & 5 \\ %
    6  & \ind \ind \ind  \textbf{\color{DarkOrange}return} {\color{DarkBlue}y}
    & & \*  & \-  & \-  & \-  & \-  & \*  & 6 \\ %
    7  & \ind  \textbf{\color{DarkOrange}else}:
    & & \-  & \-  & \*  & \*  & \-  & \-  & 7 \\ %
    8 & \ind \ind  \textbf{\color{DarkOrange}if} {\color{DarkBlue}x} > {\color{DarkBlue}y}:
    & & \-  & \-  & \*  & \*  & \-  & \-  & 8 \\ %
    9 & \ind \ind \ind  \textbf{\color{DarkOrange}return} {\color{DarkBlue}y}
    & & \-  & \-  & \*  & \-  & \-  & \-  & 9 \\ %
    10 & \ind \ind  \textbf{\color{DarkOrange}elif} {\color{DarkBlue}x} > {\color{DarkBlue}z}:
    & & \-  & \-  & \-  & \*  & \-  & \-  & 10 \\ %
    11 & \ind \ind \ind  \textbf{\color{DarkOrange}return} {\color{DarkBlue}x}
    & & \-  & \-  & \-  & \*  & \-  & \-  & 11 \\ %
    12 & \ind  \textbf{\color{DarkOrange}return} {\color{DarkBlue}z}
    & & \-  & \-  & \-  & \-  & \*  & \-  & 12 \\ %
       &
    & & \PASS{} & \PASS{} & \PASS{} & \PASS{} & \PASS{} & \FAIL{}\\
    \end{tabular}
    \caption{Statistical fault localization~\cite{jones2002tarantula}.
    The \texttt{middle()} function takes three values and returns the one that is neither the smallest nor the largest.
    Execution of Line~6 correlates most closely with the failure.}%
    \label{fig:middle-example}
    \Description{A table with the twelve lines of the Python function middle(x, y, z) as rows and six test runs as columns. Each column gives the inputs x, y, and z, marks the lines the run covers, and ends with its outcome. Five runs pass, and the last one, with x = 2, y = 1, and z = 3, fails. Line 6, which returns y, is covered by the failing run and by one passing run.}
\end{wrapfigure}

Repair agents built on large language models (LLMs) have changed what automated program repair can do.
Given an issue report and a repository, agents such as \SWEAGENT{}~\cite{yang2024sweagent}, \AUTOCODEROVER{}~\cite{zhang2024autocoderover}, or \AGENTLESS{}~\cite{xia2025agentless} explore the code, edit it, run it, and submit a patch.
They resolve a substantial share of real GitHub issues of benchmarks such as \SWEBENCH{}~\cite{jimenez2024swebench}.
Before an agent can fix a failure, however, it has to understand it.
All it starts from is the issue text, a user's description of a symptom that may or may not hint at the cause.
Much of an agent's effort thus goes into reconstructing what goes wrong, by reading code, adding print statements, and writing reproduction scripts.
When that reconstruction fails, so does the repair.
The classic aid for this reconstruction is \emph{fault localization}.
\Cref{fig:middle-example} shows its best-known form, statistical fault localization~\cite{jones2002tarantula, abreu2007accuracy}.
It correlates the lines that tests execute with their outcomes and ranks Line~6 of the \texttt{middle()} function as most suspicious.
Some agents receive such a ranking~\cite{zhang2024autocoderover, bouzenia2025repairagent}.
The ranking is valuable, but it only tells \emph{where} to look, not \emph{why} the program fails.
Line~6 is also executed by a passing test.
What distinguishes the failing run is that it reaches Line~6 with \texttt{y} smaller than \texttt{x}.
The passing run reaches it with \texttt{x} equal to \texttt{y}, and returning \texttt{y} is only correct in the latter case.
Developers rarely profit from a ranking of locations alone~\cite{parnin2011automated}.
What they need, and build when they debug, is an account of how the failure comes about.
This account is a \emph{cause-effect chain} from the circumstances of the failing run to the failure~\cite{zeller2002causeeffect}.
Such failure explanations are largely missing between debugging and repair~\cite{smytzek2026diagnosis}.

In this paper, we present \FORECAST{}, which provides repair agents with such an account before they start.
\FORECAST{} runs the failing tests and then the passing tests that share covered code with them.
It records fine-grained \emph{execution features}~\cite{smytzek2026efdd}, such as which branches are taken, which conditions hold, how values compare, where variables are defined and used, and which functions are called.
It merges all runs into a \emph{traced relevance tree}, a call tree whose nodes collect the features observed in passing and in failing runs.
At every function where the two differ, it selects the feature, or the conjunction or disjunction of features, that separates them most sharply.
Read in the order of execution and simplified, these features form a \emph{cause chain}.
A cause chain is a short sequence of conditions, data flows, values, and calls that mark where the failing run departs from the passing ones on its way to the failure.
Each cause comes with the file and line where it was observed and a \emph{discriminability} that quantifies how sharply it separates failing from passing runs.
We speak of causes because each of them is verified empirically to separate the failing from the passing runs.
We speak of a chain because the causes are ordered by execution, not because \FORECAST{} proves that one causes the next.
Whether removing a cause would prevent the failure could only be established by experiments that alter the program state~\cite{zeller2002causeeffect}, which \FORECAST{} does not perform.
The cause chain instead provides evidence of where and how the failing run departs from the passing ones, from which an agent can form and check its own hypotheses.
\FORECAST{} appends this \emph{cause-chain diagnosis} of the failure to the issue description an agent receives.
The agent itself is not modified.
Whatever the agent does with the issue, it now does with the diagnosis, which makes \FORECAST{} applicable to any agent that builds its prompts from an issue.
Computing a cause-chain diagnosis requires no LLM.
Once computed, the diagnosis can serve any number of repair attempts, agents, and models.
\FORECAST{} considers a test that reproduces the failure as given, for instance, developer-written tests, failing builds, and bug reports that ship with a test.
Obtaining such a test is left to other work.

We evaluate \FORECAST{} on all $\nInstances$ issues of \SWELITE{}~\cite{jimenez2024swebench} with five repair agents of different designs (\MINISWEAGENT{}, \SWEAGENT{}, \AUTOCODEROVER{}, \REPAIRAGENT{}, and \AGENTLESS{}).
Each agent runs with the commercial \HAIKU{} and with the open-weight \QWEN{}, once with the cause-chain diagnosis and once without it, which makes $\nAttempts$ repair attempts in total.
We derive the diagnoses from the benchmark's developer tests.

Our evaluation yields four main results.
First, the cause-chain diagnosis makes every agent better.
All five agents resolve significantly more issues with the diagnosis than without it, with either model.
Over all agents, the share of resolved issues rises from $\haikuBase$ to $\haikuForecast$ with \HAIKU{} and from $\qwenBase$ to $\qwenForecast$ with \QWEN{}.
This lifts the open-weight model above the commercial model without the diagnosis.
Even \REPAIRAGENT{}, which already runs the very tests the diagnosis is derived from, gains more than any other agent.
Second, the gains are robust.
The agents do lose some issues they resolve without the diagnosis, but for each of them, they gain about $\haikuGainsPerLoss$ others with \HAIKU{} and $\qwenGainsPerLoss$ with \QWEN{}.
When we repeat the runs, every run with the diagnosis resolves significantly more issues than every run without it.
Third, the gains stem from what the cause chain says about the failure at hand.
Given the cause chain of another issue, the agents resolve about as many issues as without any diagnosis, so an unrelated diagnosis neither helps nor causes detectable harm.
With the causes in random order or reduced to their locations, the agents lose most of the gain.
Finally, the cause-chain diagnosis keeps fixes small and saves LLM tokens.
The changes to existing code remain as small as without the diagnosis.
The tokens spent per resolved issue fall by $\haikuTokenDrop$ with \HAIKU{} and $\qwenTokenDrop$ with \QWEN{}.
Computing a diagnosis needs no LLM and takes a median of $\diagMedianMinutes$ minutes of machine time per issue.

Overall, we make the following contributions:
\begin{description}[leftmargin=5pt,noitemsep,partopsep=0pt,topsep=0pt,parsep=0pt]
 \item[A novel \emph{approach for inferring cause-chain diagnoses}.] We introduce a technique that infers a cause-chain diagnosis from passing and failing test executions.
 It merges all runs into a traced relevance tree and compares the failing runs with the passing runs at every call.
 It then combines the features that separate the failing from the passing runs into conjunctions and disjunctions and adds the calls that only failing runs make.
 Finally, it orders the resulting causes as they occur in the execution.
 The resulting cause chain tells where in the execution, and in which order, the failing run departs from the passing ones.
 \item[A way to \emph{enhance repair agents} without modifying them.] \FORECAST{} appends the cause-chain diagnosis to the issue an agent receives.
 The diagnosis thus reaches every agent that builds its prompts from the issue, whatever the agent's actual design is.
 \item[An extensive \emph{evaluation} of \FORECAST{}.] We enhance five repair agents of different designs with our cause-chain diagnoses and evaluate them on all $\nInstances$ issues of \SWELITE{} with a commercial and an open-weight model, $\nAttempts$ repair attempts in total.
 We repeat the runs to measure how much they vary, and we run thirteen variants of the diagnosis that withhold or substitute its parts.
 \item[A \emph{replication package}.] \FORECAST{}, the evaluation scripts, and every diagnosis and repair run are publicly available~\cite{forecast-artifact}.
\end{description}

The remainder of this paper is structured as follows.
\Cref{sec:repair-agents} introduces the repair agents we enhance.
\Cref{sec:execution-features} explains execution features and their role in fault localization and debugging, as introduced by our prior work on \EFDD{}~\cite{smytzek2026efdd}.
\Cref{sec:debugging} presents our approach to infer cause-chain diagnoses from execution features.
\Cref{sec:enhancing-repair-agents} describes how we enhance repair agents with these diagnoses.
\Cref{sec:implementation} details the implementation of our approach.
\Cref{sec:evaluation} evaluates how the diagnoses change the effectiveness, fix quality, and cost of five repair agents, how repeatable these changes are, and which parts of the diagnosis they stem from.
We address potential threats to validity in \Cref{sec:threats-to-validity} and review related work in \Cref{sec:related-work}.
Finally, \Cref{sec:conclusion} concludes the paper and suggests directions for future research.

\section{Repair Agents}%
\label{sec:repair-agents}

Automated program repair aims to reduce the manual effort required to fix bugs in software systems.
Traditional repair approaches relied on pre-defined templates or mutation operators to generate patches~\cite{legoues2019automated, monperrus2018automatic}.
Large language models (LLMs) have since enabled a new generation of repair agents.
These agents understand bug reports, navigate codebases, and generate fixes autonomously, without domain-specific knowledge or hand-crafted repair templates.

A repair agent has to understand the bug report or issue description, localize the fault in the codebase, generate candidate patches, and validate them by running tests or other checks.
The quality of its output depends on how accurately it understands the failure and locates the root cause of the bug.
Its effectiveness is therefore often limited by the quality of information it receives about the failure.
Vague bug reports or insufficient context can lead to incorrect or overfitted patches.
Such patches pass the immediate tests but fail to address the underlying issue or introduce new bugs.

In this work, we enhance repair agents with the cause-chain diagnoses of \FORECAST{}.
These diagnoses describe the chain of causes that leads to a failure, so that an agent does not have to reconstruct it from the issue alone.
We focus on five established repair agents of different designs.

\paragraph{\AUTOCODEROVER{}.}
\AUTOCODEROVER{}~\cite{zhang2024autocoderover} is an autonomous program repair agent that combines code search with LLM-based patch generation.
It operates in two main phases.
In a context retrieval phase, it uses a combination of static analysis and search techniques to identify relevant code elements.
In a repair phase, it generates and validates patches.
\AUTOCODEROVER{} navigates large codebases through search APIs over the program structure, which retrieve classes, methods, and code snippets by name.
The agent iteratively refines its understanding by analyzing class definitions, method signatures, and their relationships.
It uses this knowledge to guide its search for bug-related code.
Once relevant code locations are identified, \AUTOCODEROVER{} uses an LLM to generate patches that are then validated against the test suite.
Its search is driven by the names of classes and methods.
A cause-chain diagnosis names the functions and lines where the failing run departs from the passing ones and thus gives it concrete starting points.

\paragraph{\AGENTLESS{}.}
\AGENTLESS{}~\cite{xia2025agentless} takes a lightweight, three-phase approach to program repair that achieves competitive results with minimal complexity.
Agent-based systems engage in extended interactions with the environment.
\AGENTLESS{} instead performs localization, repair, and patch validation in a fixed pipeline.
In the localization phase, it narrows the repository down hierarchically, from suspicious files to classes and functions and finally to the lines to edit.
To do so, it prompts the LLM with the repository structure and the code, combined with embedding-based retrieval of files.
The repair phase then samples multiple candidate patches for these locations in a simple search-and-replace format.
Finally, the validation phase runs regression tests and reproduction tests that \AGENTLESS{} generates itself, and selects among the remaining candidates by majority vote.
With this design, \AGENTLESS{} resolved about 27\% of \SWELITE{} at a fraction of the cost of contemporary agents, while its steps remain easy to interpret.
Each of its phases starts from the issue.
A cause-chain diagnosis appended to the issue thus reaches localization, repair, and the generation of reproduction tests alike.
Since \AGENTLESS{} already generates and runs a test that reproduces the issue, it also lets us study whether a diagnosis adds value beyond what such a test provides.

\paragraph{\SWEAGENT{}.}
\SWEAGENT{}~\cite{yang2024sweagent} introduces the concept of Agent-Computer Interfaces (ACI) for the interaction between LLM-based agents and software engineering environments.
Rather than providing the agent with raw shell access, \SWEAGENT{} designs specialized commands and feedback mechanisms that suit the agent's capabilities and limitations.
For example, it includes custom file viewing commands that provide line numbers and pagination, search functionality tailored to code navigation, and edit commands with built-in validation.
The ACI design allows \SWEAGENT{} to work more effectively within the constraints of LLM context windows and reasoning capabilities.
\SWEAGENT{} operates in an interactive loop.
It reads the issue description, explores the codebase using its specialized commands, formulates hypotheses about the bug location, generates edits, and validates them by running tests.
Much of this loop is spent on finding where to look, which is what a cause-chain diagnosis can shorten.

\paragraph{\MINISWEAGENT{}.}
\MINISWEAGENT{}~\cite{sweagent2025minisweagent}, by the authors of \SWEAGENT{}, takes the opposite stance on the agent-computer interface and deliberately provides none.
The LLM interacts with the repository through shell commands alone, and each command runs independently of the previous ones.
The whole interaction is kept as a single linear history of messages.
The agent explores the codebase, edits files, and runs tests with ordinary command-line tools until it submits its changes as a patch.
Because it adds no specialized tools or heuristics of its own, \MINISWEAGENT{} shows how much a repair depends on the information the LLM receives.
It is thus a direct probe of what a cause-chain diagnosis contributes.

\paragraph{\REPAIRAGENT{}.}
\REPAIRAGENT{}~\cite{bouzenia2025repairagent} is an autonomous repair agent that treats the LLM as an agent that plans and executes its own actions.
It acts through a set of repair tools for reading and searching code, extracting tests, running them, and writing and trying candidate fixes.
Its prompt is updated dynamically with the outcome of every action.
A finite state machine guides which tools the agent may use as it moves between understanding the bug, collecting the information needed to fix it, and trying out fixes.
Throughout, the agent states and revises explicit hypotheses about the bug.
\REPAIRAGENT{} was evaluated on the \DEFECTS4J{} Java benchmark, where it fixed 164 bugs, including 39 that prior techniques had not fixed.
It already runs and inspects the test suite as part of its repair loop.
\REPAIRAGENT{} thus lets us study whether a cause-chain diagnosis adds value beyond the information an agent can extract from the tests itself.

Despite their different designs, all five agents have to identify the root cause of a failure from limited information.
By enhancing these agents with the cause-chain diagnoses of \FORECAST{}, we aim to improve their ability to generate correct fixes and to reduce the exploration they need to get there.

\section{Execution Features}%
\label{sec:execution-features}

Execution features are observable properties of program behavior, captured during test execution, that characterize how the program behaves under different inputs.
Unlike static code analysis, they describe what actually happens when the program runs, which provides concrete evidence that can distinguish between executions.

In fault localization and debugging, execution features are the foundation for understanding why some tests pass while others fail.
By comparing the execution features observed in passing and in failing tests, we can identify the behaviors that are unique to failures.
These behaviors pinpoint the conditions under which bugs manifest.
This discriminative approach has proven effective in automated debugging~\cite{smytzek2026efdd}, as it captures detailed program state and control flow information beyond code coverage.
\FORECAST{} uses a catalog of execution features organized into several categories by the type of program behavior they capture.

\paragraph{Structural Features.}
These features describe which parts of the code are executed.
Line execution features record whether specific source code lines are covered during a test run, and function call features track which functions are invoked.
Loop features capture whether a loop is executed and also how often it iterates (never, once, or multiple times).

\paragraph{Control Flow Features.}
These features capture decisions made during program execution.
Branch features record which branches of conditional statements are taken (then-branch versus else-branch).
Condition features capture the Boolean evaluation of specific conditions in the code.
For example, a condition feature might record that \texttt{x > 0} evaluated to \textsc{True} at a particular line.

\paragraph{Data Flow Features.}
These features track how data moves through the program.
Def-use features capture relationships between variable definitions and their subsequent uses, recording where a variable was assigned and where that value was later accessed.
Return value features monitor what functions return, including comparisons against zero or \texttt{None}.
Variable predicate features evaluate properties of variables such as whether they are zero, positive, negative, or \texttt{None}.
Scalar pair features compare two variables or a variable against a predefined value, recording relationships like equality, ordering, or inequality.

\paragraph{Property Features.}
These features evaluate higher-level properties of program values.
String property features check characteristics like whether a string is ASCII-only, contains digits, contains special characters, or is empty.
Length features measure the length of collections and strings, categorizing them as empty, singleton, or multi-element.
These features may seem less directly actionable than structural or control flow features.
They can, however, identify symptomatic behaviors that help characterize bug manifestations, particularly in programs that process text or collections.

Execution features are useful because they can discriminate between outcomes.
A feature is discriminative if it appears in both passing and failing tests but with different values or frequencies.
For example, a condition \texttt{token.isnumeric()} may evaluate to \textsc{True} in all passing tests but to \textsc{False} in all failing tests.
This feature then perfectly discriminates between the two outcomes.
Such discriminative features characterize how failing executions differ from passing ones, which often points towards the root cause.
However, it is worth noting that a feature can also be a consequence of the fault rather than its origin.

\FORECAST{} builds upon the execution feature infrastructure of our prior work on Execution Feature Driven Debugging (\EFDD{})~\cite{smytzek2026efdd}.
\EFDD{} showed that fault localization benefits from execution features beyond line coverage.
It learns a decision tree that separates passing from failing runs by the features each run exhibits as a whole.
\FORECAST{} instead compares the runs at every call of a traced relevance tree.
Its causes thus say not only which features discriminate, but also where and in which order they occur on the way to the failure.

\section{Inferring Cause-Chain Diagnoses}%
\label{sec:debugging}

\begin{figure}[t]
  \centering
  \begin{tikzpicture}[
    sn/.style={circle, draw, inner sep=0pt, minimum size=3.5mm,
               font=\tiny},
    diag/.style={draw, rounded corners=2pt, fill=forecastblue!18,
                 minimum height=9mm, minimum width=1.75cm,
                 align=center, font=\footnotesize},
    artifact/.style={font=\footnotesize\itshape, align=center},
    data/.style={below, font=\scriptsize\itshape},
    arr/.style={-{Stealth[length=2mm]}, thick},
  ]
    \node[artifact] (input) {Program\\[-2pt]and Tests};
    \node[sn, above=1pt of input] (s1) {1};

    \node[diag, right=0.7cm of input] (collect) {Instrument\\[-1pt]and Trace};
    \node[sn, above=1pt of collect] (s2) {2};
    \node[diag, right=0.75cm of collect] (tree) {Build\\[-1pt]Tree};
    \node[sn, above=1pt of tree] (s3) {3};
    \node[diag, right=0.75cm of tree] (extract) {Extract\\[-1pt]Features};
    \node[sn, above=1pt of extract] (s4) {4};
    \node[diag, right=0.75cm of extract] (combine) {Combine\\[-1pt]Atoms};
    \node[sn, above=1pt of combine] (s5) {5};
    \node[diag, right=0.75cm of combine] (simplify) {Simplify\\[-1pt]Chain};
    \node[sn, above=1pt of simplify] (s6) {6};

    \node[artifact, right=0.7cm of simplify] (chain) {Cause\\[-2pt]Chain};
    \node[sn, above=1pt of chain] (s7) {7};

    \draw[arr] (input) -- (collect);
    \draw[arr] (collect) -- node[data] {traces} (tree);
    \draw[arr] (tree) -- node[data] {tree} (extract);
    \draw[arr] (extract) -- node[data] {atoms} (combine);
    \draw[arr] (combine) -- node[data] {causes} (simplify);
    \draw[arr] (simplify) -- (chain);

    \node[draw, thick, rounded corners=4pt, inner xsep=6pt, inner ysep=5pt,
          fit=(collect) (simplify) (s2)] (forecast) {};
    \node[font=\footnotesize\bfseries, anchor=south west]
      at ([yshift=1pt]forecast.north west) {\FORECAST{}};
  \end{tikzpicture}
  \caption{Diagnosis pipeline of \FORECAST{}.
    Given a program and its failing and passing tests~\circnum{1}, \FORECAST{} instruments the program and traces the tests~\circnum{2} (\Cref{sec:implementation}),
    builds a traced relevance tree from the traces~\circnum{3} (\Cref{sub:traced-relevance-trees}),
    extracts the features that separate the failing from the passing runs at its comparable nodes~\circnum{4} (\Cref{sub:features}),
    combines them into conjunctions and disjunctions~\circnum{5} (\Cref{sub:combining}),
    and simplifies the result~\circnum{6} (\Cref{sub:simplifying}) into a cause chain~\circnum{7}.}%
  \label{fig:diagnosis-pipeline}
  \Description{A flow diagram of seven numbered steps from left to right. The program and its tests (1) enter FORECAST, which instruments and traces them (2), builds a tree from the traces (3), extracts features from the tree as atoms (4), combines the atoms into causes (5), and simplifies the causes (6) into the cause chain (7).}
\end{figure}
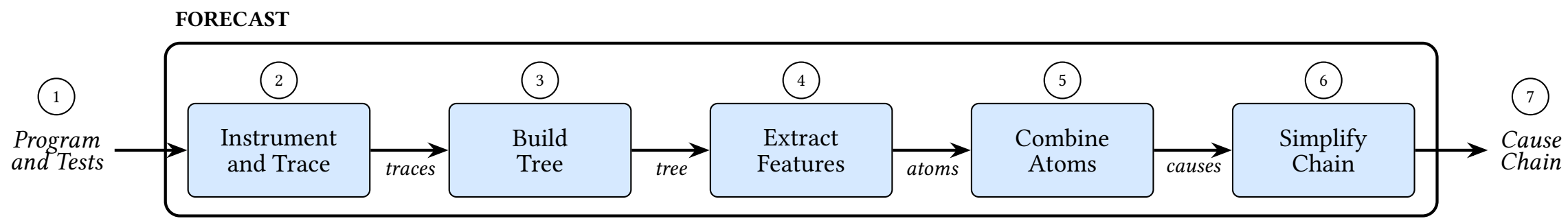

The core of \FORECAST{} infers a cause-chain diagnosis from the executions of passing and failing tests.
Its cause chain consists of execution features that separate the failing from the passing runs, ordered as they occur on the way from the start of a test to its failure.
Fault localization techniques identify suspicious code locations.
A cause chain instead describes \emph{how} the failing execution departs from the passing ones.
Its causes are verified empirically, as observations that separate the failing from the passing runs.
That they cause one another is not verified, so a cause chain is evidence for an agent to act upon, not a proof of why the program fails.
This section describes how \FORECAST{} infers cause-chain diagnoses in four stages.
It builds traced relevance trees, extracts discriminative features, combines atomic features into complex predicates, and simplifies the resulting cause chain for actionability.
\Cref{fig:diagnosis-pipeline} gives an overview of these stages and of the collection of traces they start from.

\subsection{Running Example}

Throughout this section, we use a simple expression parser as a running example to illustrate how \FORECAST{} infers a cause-chain diagnosis.
The parser evaluates arithmetic expressions but contains a subtle bug in the subtraction operator (\Cref{fig:expr-parser-bug}).

\begin{figure}[t]
\centering
\def\ind{\qquad}

\begin{subfigure}[t]{0.48\textwidth}
\centering
\scriptsize
\begin{tabular}{rl}
\multicolumn{2}{l}{\texttt{expr/arithmetic.py}} \\
   & \texttt{\ind \ldots} \\
6  & \texttt{\textbf{\color{DarkOrange}class} {\color{Blue}Binary}({\color{Blue}Term}):} \\
7  & \texttt{\ind \textbf{\color{DarkOrange}def} {\color{Blue}\_\_init\_\_}({\color{DarkBlue}self}, {\color{DarkBlue}left}, {\color{DarkBlue}right}):} \\
\rowcolor{yellow!20}
8  & \texttt{\ind \ind {\color{DarkBlue}self}.{\color{Blue}left} = {\color{DarkBlue}left}} \\
9  & \texttt{\ind \ind {\color{DarkBlue}self}.{\color{Blue}right} = {\color{DarkBlue}right}} \\
   & \texttt{\ind \ldots} \\
17 & \texttt{\textbf{\color{DarkOrange}class} {\color{Blue}Sub}({\color{Blue}Binary}):} \\
\rowcolor{yellow!20}
18 & \texttt{\ind \textbf{\color{DarkOrange}def} {\color{Blue}evaluate}({\color{DarkBlue}self}):} \\
\rowcolor{red!15}
19 & \texttt{\ind \ind \textbf{\color{DarkOrange}return} {\color{DarkBlue}self}.{\color{Blue}left}.{\color{Blue}evaluate}() {\color{red}+} {\color{DarkBlue}self}.{\color{Blue}right}.{\color{Blue}evaluate}()} \\
   & \texttt{\ind \ind {\color{gray}\# Bug: should be -}} \\
   & \texttt{\ind \ldots} \\
22 & \texttt{\textbf{\color{DarkOrange}class} {\color{Blue}Mul}({\color{Blue}Binary}):} \\
\rowcolor{yellow!20}
23 & \texttt{\ind \textbf{\color{DarkOrange}def} {\color{Blue}evaluate}({\color{DarkBlue}self}):} \\
\rowcolor{yellow!20}
24 & \texttt{\ind \ind \textbf{\color{DarkOrange}return} {\color{DarkBlue}self}.{\color{Blue}left}.{\color{Blue}evaluate}() * {\color{DarkBlue}self}.{\color{Blue}right}.{\color{Blue}evaluate}()} \\
   & \texttt{\ind \ldots} \\
\end{tabular}
\caption{Buggy \texttt{Sub} class}%
\label{fig:expr-arithmetic}
\end{subfigure}
\hfill
\begin{subfigure}[t]{0.48\textwidth}
\centering
\scriptsize
\begin{tabular}{rl}
\multicolumn{2}{l}{\texttt{expr/parse.py}} \\
   & \texttt{\ind \ldots} \\
15 & \texttt{\textbf{\color{DarkOrange}def} {\color{Blue}parse}({\color{DarkBlue}s}):} \\
   & \texttt{\ind \ldots} \\
21 & \texttt{\ind {\color{DarkBlue}tokens} = {\color{Blue}list}({\color{Blue}reversed}({\color{DarkBlue}s}.{\color{Blue}split}({\color{red}" "})))} \\
   & \texttt{\ind \ldots} \\
23 & \texttt{\ind {\color{DarkBlue}term} = {\color{Blue}parse\_add\_sub}({\color{DarkBlue}tokens})} \\
   & \texttt{\ind \ldots} \\
62 & \texttt{\textbf{\color{DarkOrange}def} {\color{Blue}parse\_add\_sub}({\color{DarkBlue}tokens}):} \\
63 & \texttt{\ind {\color{DarkBlue}term} = {\color{Blue}parse\_mul\_div}({\color{DarkBlue}tokens})} \\
64 & \texttt{\ind \textbf{\color{DarkOrange}if} {\color{DarkBlue}tokens} \textbf{\color{DarkOrange}and} {\color{DarkBlue}tokens}[0] \textbf{\color{DarkOrange}in} {\color{red}"+-"}:} \\
65 & \texttt{\ind \ind {\color{DarkBlue}token} = {\color{DarkBlue}tokens}.{\color{Blue}pop}(0)} \\
\rowcolor{yellow!20}
66 & \texttt{\ind \ind \textbf{\color{DarkOrange}if} {\color{DarkBlue}token} == {\color{red}"+"}:} \\
67 & \texttt{\ind \ind \ind \textbf{\color{DarkOrange}return} {\color{Blue}Add}({\color{Blue}parse\_add\_sub}({\color{DarkBlue}tokens}), {\color{DarkBlue}term})} \\
68 & \texttt{\ind \ind \textbf{\color{DarkOrange}else}:} \\
\rowcolor{yellow!20}
69 & \texttt{\ind \ind \ind \textbf{\color{DarkOrange}return} {\color{Blue}Sub}({\color{Blue}parse\_add\_sub}({\color{DarkBlue}tokens}), {\color{DarkBlue}term})} \\
   & \texttt{\ind \ldots} \\
\end{tabular}
\caption{Parser creating \texttt{Sub} object}%
\label{fig:expr-parse}
\end{subfigure}

\vspace{0.2cm}

\begin{subfigure}[t]{\textwidth}
\centering
\scriptsize
\setlength{\tabcolsep}{4pt}
\begin{tabular}{lccccccccc}
\toprule
\textbf{Input} & \texttt{2 + 3} & \texttt{5 / 6} & \texttt{1 + (2 * 3)} & \texttt{(1 + 2) * 3} & \texttt{1 + 1} & \texttt{1 - 0} & \texttt{\textasciitilde{} 4} & \cellcolor{red!10}\texttt{2 - 3} & \cellcolor{red!10}\texttt{2 * (1 - 1)} \\
\midrule
\textbf{Expected} & 5 & 0.833... & 7 & 9 & 2 & 1 & -4 & \cellcolor{red!10}-1 & \cellcolor{red!10}0 \\
\textbf{Actual} & 5 & 0.833... & 7 & 9 & 2 & 1 & -4 & \cellcolor{red!10}\textbf{5} & \cellcolor{red!10}\textbf{4} \\
\textbf{Result} & \PASS{} & \PASS{} & \PASS{} & \PASS{} & \PASS{} & \PASS{} & \PASS{} & \cellcolor{red!10}\FAIL{} & \cellcolor{red!10}\FAIL{} \\
\bottomrule
\end{tabular}
\caption{Test suite (7 passing, 2 failing)}
\label{fig:expr-tests}
\end{subfigure}

\caption{Expression parser with subtraction bug.
Line~19 in \texttt{arithmetic.py} incorrectly uses addition (\texttt{+}) instead of subtraction (\texttt{-}).
\texttt{parse()} reverses the tokens (Line~21), so the parser reads an expression from right to left.
Yellow highlights indicate lines referenced in the final diagnosis.
Red highlights show the bug location and failing tests.}
\label{fig:expr-parser-bug}
    \Description{Three panels. The first shows the classes Binary, Sub, and Mul of expr/arithmetic.py, where Line 19 of Sub.evaluate adds the two operands instead of subtracting them. The second shows the functions parse and parse_add_sub of expr/parse.py, where Line 69 creates a Sub object. The third is a table of nine tests with their input, expected, and actual result. Seven tests pass, and the tests 2 - 3 and 2 * (1 - 1) fail, returning 5 and 4 instead of -1 and 0.}
\end{figure}

The bug on Line~19 causes the \texttt{Sub} class to add instead of subtract, so that expressions like \texttt{2 - 3} evaluate to 5 instead of -1.
\FORECAST{} analyzes the execution traces of passing tests (like \texttt{2 + 3}) and failing tests (like \texttt{2 - 3}).
From these traces, it automatically infers a cause-chain diagnosis of this failure.

\subsection{Traced Relevance Trees}%
\label{sub:traced-relevance-trees}

To compare passing and failing runs at every call of the program, \FORECAST{} constructs a \emph{traced relevance tree}, a unified call tree across all test executions.
A call tree usually describes a single execution.
The traced relevance tree instead merges the executions of passing and failing tests into a single structure, so that their paths can be compared directly.

\subsubsection{Tree Construction}

Each node in the traced relevance tree represents a function together with the sequence of calls that leads to it from the root.
The children of a node are the functions it calls.
A function that is reached along two different call paths thus has two nodes.
The tree unifies the traces of all test runs.
Runs that follow the same call path share its nodes, while divergent paths create separate branches.
For each function invocation, the tree tracks two sets of feature vectors:

\begin{description}
    \item[Enter features ($\phi_{\text{enter}}$):] Features observed in the caller since its previous call boundary, recorded when the function is entered, that is, what leads up to the call
    \item[Exit features ($\phi_{\text{exit}}$):] Features observed in the function since its last nested call returned (or since its entry), recorded when it returns, that is, what the function does before it returns
\end{description}

With these two sets, \FORECAST{} observes both what leads into a function and what the function does before it returns, either of which may contribute to a failure.
Each feature vector is tagged with its test result (\textsc{Passing} or \textsc{Failing}), which the subsequent discriminative analysis needs.

\begin{algorithm}[t]
\caption{Building Traced Relevance Trees}%
\label{alg:build-tree}
\begin{algorithmic}[1]
\Function{BuildTree}{$failing, passing$}
    \State $root \gets$ empty tree with root node
    \For{each $execution$ in $failing$, then in $passing$}
        \State $current \gets root$
        \State $features \gets$ empty feature vector (tagged with test result)
        \For{each $event \in execution$}
            \State extract features from $event$ into $features$
            \If{$event$ is function entry}
                \If{$execution$ is passing and $current$ has no child for this function}
                    \State skip the events up to the matching exit \Comment{no failing run reaches it}
                    \State $features \gets$ empty feature vector (tagged with test result) \Comment{discards them}
                    \State \textbf{continue}
                \ElsIf{the function is not $current$'s own} \Comment{direct recursion stays at $current$}
                    \State $current \gets$ child node for this function, created if missing
                \EndIf
                \State \Call{Keep}{$current.enter\_features$, $execution$, $features$}
                \State $features \gets$ empty feature vector (tagged with test result)
            \ElsIf{$event$ is function exit}
                \State \Call{Keep}{$current.exit\_features$, $execution$, $features$}
                \State $features \gets$ empty feature vector (tagged with test result)
                \State $current \gets$ parent node, unless a direct recursive call returns
            \EndIf
        \EndFor
    \EndFor
    \State \Return $root$
\EndFunction
\Statex
\Function{Keep}{$vectors, execution, v$} \Comment{at most two vectors per run and node}
    \If{$vectors$ holds fewer than two vectors of $execution$}
        \State append $v$ to $vectors$
    \Else
        \State replace the later vector of $execution$ in $vectors$ by $v$ \Comment{first and most recent call}
    \EndIf
\EndFunction
\end{algorithmic}
\end{algorithm}

\Cref{alg:build-tree} illustrates the tree construction process.
It processes the failing executions first, so that they span the tree.
It then folds in the passing executions only along the call paths the failing ones take (\Cref{sec:implementation}), since a node that no failing run reaches cannot be compared.
The algorithm collects the features of all other events, such as branches, conditions, or variable updates, in the current vector.
At every call boundary, it saves that vector to the entered node's enter features or the exited node's exit features and starts a fresh one.
Each vector thus holds only the \emph{local} features observed between two call boundaries.
A passing run may call a function that no failing run reaches.
It then records neither the features inside that call nor those collected before it, which would be the entry features of a node that does not exist.
Once the call returns, it continues with a fresh vector.
A caller's features are recorded and compared at the caller's own node.
The call tree already encodes the path that leads to a node, so carrying these features into every descendant would only duplicate them.
A directly recursive call stays at its caller's node, so that a function calling itself does not deepen the tree.
A function that is called again through other functions gets a node of its own.

A function that a single run calls many times would contribute one entry and one exit vector per call.
Hot code in one run could then outweigh all other runs.
\FORECAST{} therefore keeps, per node and run, at most two entry and two exit vectors, namely those of the first call and of the most recent one, which each further call replaces.
Every run thus stays represented at every node it reaches, with a bounded weight.
The counts that enter the discriminability score (\Cref{sub:features}) are therefore counts of such observations, at most two per run and node, rather than of tests.

\subsubsection{Comparable Node Marking}

Not all nodes in the traced relevance tree contribute to distinguishing between passing and failing tests.
\FORECAST{} identifies \emph{comparable nodes}, the nodes at which passing and failing runs can be compared, together with the calls that lead to them.
The marking proceeds recursively and bottom-up, so that only relevant execution paths are analyzed.

A node is marked as comparable if it exhibits \emph{local comparability} or \emph{propagated comparability}.
A node is locally comparable when at least one feature is observed in \emph{both} passing and failing test vectors at that node, in either its enter or its exit features.
The values of this feature in the two groups can then be compared.
For example, a condition \texttt{token.isnumeric()} may evaluate to \textsc{True} in some passing tests but to \textsc{False} in failing tests.
The node then exhibits local comparability.
A node has propagated comparability when any of its child nodes is comparable, which indicates that the failure manifests in a deeper call.
This preserves the entire call chain that leads to comparable functions in the analysis.

The algorithm first recursively processes all children to mark comparable nodes in the subtrees and then evaluates the current node's local comparability.
For this local evaluation, it collects the features that appear in the passing and in the failing enter vectors at this node, and likewise for the exit vectors.
If the passing and failing features of the enter or of the exit vectors intersect, passing and failing runs can be compared at this node, and it is marked comparable.
Whether a feature actually separates the runs is decided only when it is scored (\Cref{sub:features}).
A node whose features take the same values in both groups contributes no cause.

The recursive marking thus focuses \FORECAST{} on the execution paths relevant to the failure.
Functions that only passing or only failing runs reach are excluded, because there is nothing to compare them against.
Calls that only failing runs make are reported as divergences instead (\Cref{sub:features}).
The bottom-up propagation preserves the entire call chain that leads to a comparable function.
The cause chain can thus follow the calls from the start of the test to the failure.

\subsection{Extracting Relevant Features}%
\label{sub:features}

Once the comparable nodes are identified, \FORECAST{} extracts the features that best separate passing from failing tests.
It does so independently at each comparable node and examines both enter and exit features to capture state changes throughout the execution path.

\subsubsection{Cause Chain Extraction}

\FORECAST{} traverses the traced relevance tree in depth-first order and extracts discriminative features at each comparable node, which yields a cause chain in the order of execution.
At each comparable node, \FORECAST{} first extracts the most discriminative cause from the enter features, which describe what leads up to the call.
Next, it recursively processes all comparable child nodes to capture deeper execution contexts where the failure may further manifest.
Finally, it extracts the most discriminative cause from the exit features, which describe what the function does before it returns.

This ordering of enter features, children, and exit features follows the order of execution along every call path.
What leads into a function comes first, followed by the causes from nested function calls, and finally what the function does before it returns.
Since the tree is built from the failing runs first, the children of a node are visited in the order in which the failing runs first call them.
Sibling calls thus also appear in the order in which they occur.
Because the tree merges all runs, a cause chain is an ordered collection of observations rather than the trace of one particular run.
With several failing tests, two causes at different nodes may stem from different failing runs.
With a single failing test, as in the median instance of our evaluation, all of them stem from the same run.

Non-comparable nodes contribute no causes, so only execution features that actually differ between passing and failing tests appear in the final cause chain.
The recursive structure lets the cause chain contain both coarse-grained causes in outer functions and fine-grained causes in inner functions.
Together, they show how the failure propagates through the call stack.

\subsubsection{Feature Representation}

Features in \FORECAST{} represent observable program properties of seven kinds.
These are conditions (Boolean predicates such as \texttt{tokens[0] in '*/'}), variables (scalar comparisons such as \texttt{value != 0}), and scalar pairs (relationships between variables such as \texttt{left == right}).
They also include branches (control flow decisions), def-use relationships (data flow such as ``variable \texttt{tokens} defined at Line~62 used at Line~69''), functions (calls and returns), and lines (statement execution).
Each feature can take one of three values, \textsc{True}, \textsc{False}, or \textsc{Undefined} (if the feature was not observed during execution).
This distinguishes a condition that evaluates to false from one that is never reached.

\subsubsection{Discriminability Scoring}

To quantify how well a feature distinguishes failing from passing tests, \FORECAST{} employs the Matthews Correlation Coefficient (MCC), a balanced measure suitable for imbalanced datasets:

\begin{equation}
\text{MCC} = \frac{tp \cdot tn - fp \cdot fn}{\sqrt{(tp + fp)(tp + fn)(tn + fp)(tn + fn)}}
\end{equation}

Here, $tp$ (true positives) counts failing observations where the feature is \textsc{True}, and $fp$ (false positives) counts passing observations where the feature is \textsc{True}.
Likewise, $fn$ (false negatives) counts failing observations where the feature is not \textsc{True}, and $tn$ (true negatives) counts passing observations where the feature is not \textsc{True}.
MCC ranges from $-1$ to $+1$.
A value of $+1$ indicates perfect positive correlation (the feature is always true in failing tests and never in passing tests), $0$ indicates no correlation, and $-1$ indicates perfect negative correlation.
The counts are counts of observations, at most two per run and node (\Cref{sub:traced-relevance-trees}), rather than of tests.
The score is therefore an observation-level discriminability.
It describes how sharply a feature separates the observed failing from the observed passing executions.
It is neither a probability nor a confidence that the feature causes the failure.

For ternary features (which can be \textsc{True}, \textsc{False}, or \textsc{Undefined}), \FORECAST{} treats \textsc{Undefined} values as belonging to the ``not observed'' category.
Consider the positive hypothesis ``feature = \textsc{True} indicates failure''.
For it, $tp$ counts failing observations where the feature is \textsc{True}, and $fp$ counts passing observations where it is \textsc{True}.
Likewise, $fn$ counts failing observations where it is \textsc{False} or \textsc{Undefined}, and $tn$ counts passing observations where it is \textsc{False} or \textsc{Undefined}.
For the negative hypothesis ``feature = \textsc{False} indicates failure'', the roles are reversed, with $tp$ counting failing observations where the feature is \textsc{False}.

\FORECAST{} computes MCC scores for both the positive and negated forms of each feature and selects the form with the higher discriminability.
This bidirectional evaluation lets a feature become a cause both when it holds and when it does not hold in the failing runs.
A negated cause states that the feature was observed and evaluated to \textsc{False}, never merely that it was not observed.
For example, if failures always occur when \texttt{x == 0} evaluates to \textsc{False}, the negated form achieves a higher MCC than the positive form.

\begin{algorithm}[t]
\caption{Extracting Discriminative Features}%
\label{alg:extract-features}
\begin{algorithmic}[1]
\Function{ExtractFeatures}{$feature\_vectors$}
    \State $passing \gets \{\}$, $failing \gets \{\}$
    \State $total\_passing \gets 0$, $total\_failing \gets 0$

    \For{each $fv \in feature\_vectors$}
        \If{$fv.result = \textsc{Passing}$}
            \State $total\_passing \gets total\_passing + 1$
        \Else
            \State $total\_failing \gets total\_failing + 1$
        \EndIf
        \For{each $(feature, value) \in fv.features$}
            \State increment $passing[feature][value]$ or $failing[feature][value]$
            \State track which test vectors exhibit this $feature$ with this $value$
        \EndFor
    \EndFor

    \State $candidates \gets []$
    \For{each observed $feature$}
        \State $mcc_{pos} \gets \Call{MCC}{failing[\text{T}], passing[\text{T}], total\_failing, total\_passing}$
        \State $mcc_{neg} \gets \Call{MCC}{failing[\text{F}], passing[\text{F}], total\_failing, total\_passing}$

        \If{$mcc_{pos} \geq mcc_{neg}$}
            \State $candidates.\Call{add}{\text{Atom}(feature, mcc_{pos})}$
        \Else
            \State $candidates.\Call{add}{not \text{Atom}(feature, mcc_{neg})}$
        \EndIf
    \EndFor
    \State \Return $candidates$
\EndFunction
\end{algorithmic}
\end{algorithm}

\Cref{alg:extract-features} presents the feature extraction process.
It counts how often each feature takes each value (\textsc{True}, \textsc{False}, \textsc{Undefined}) in the passing and in the failing observations.
It also records which vectors exhibit each value, so that combinations can later be scored by set operations (intersection for AND, union for OR).
From these counts, it computes the MCC of the positive hypothesis (feature = \textsc{True} indicates failure) and of the negated one (feature = \textsc{False} indicates failure).
It keeps the hypothesis with the higher discriminability as an atomic cause, or \emph{atom}.
A feature that is \textsc{True} and \textsc{False} in the same shares of failing and passing observations cannot separate them and is skipped.

\subsubsection{Divergences}
A failing run can also leave the behavior of the passing runs altogether by calling a function that no passing run calls.
The node of such a callee holds only failing observations.
With no passing observation to compare against, none of its features can be scored, and \Cref{alg:extract-features} discards it.
However, the call itself is where the failing execution departs from the passing ones, and thus a cause candidate in its own right.
After extracting the entry cause of a comparable node, \FORECAST{} therefore inspects the node's children.
For every child that only failing runs reach, it adds a \emph{divergence} to the cause chain, stating that the failing run called $g$ from $f$, which no passing run called.
A divergence is scored with the same MCC as every other cause, for the hypothesis ``the run makes this call''.
This hypothesis holds in no passing run and in every failing run that makes the call.
A call that all failing runs make thus reaches a discriminability of 1.00.
The subsequent causes describe what follows such a call.
Ties in discriminability are therefore broken in favor of divergences when the cause chain is cut to its maximal length (\Cref{sub:simplifying}).

\subsection{Combining Atoms}%
\label{sub:combining}

Individual features often explain a failure only incompletely.
A bug may only manifest when \emph{multiple} conditions hold simultaneously (conjunction) or when \emph{any} of several conditions occurs (disjunction).
\FORECAST{} therefore systematically explores feature combinations to find more discriminative causes.

\subsubsection{Combination Strategy}

For each node in the traced relevance tree, \FORECAST{} begins with the set of atoms with the highest discriminability scores.
If these atoms achieve perfect discriminability (MCC = 1.0), no combination is necessary.
Otherwise, the algorithm explores pairwise combinations:

\begin{description}
    \item[Conjunction (AND):] Both features must be true for the combined predicate to hold
    \item[Disjunction (OR):] Either feature being true satisfies the combined predicate
\end{description}

For each combination, \FORECAST{} computes the discriminability by determining which test vectors satisfy the combined predicate.
For conjunctions, it intersects the sets of vectors where each component is true, and for disjunctions, it takes their union.

\begin{algorithm}[t]
\caption{Combining Atomic Features}%
\label{alg:combine-atoms}
\begin{algorithmic}[1]
\Function{CombineAtoms}{$atoms$}
    \State $best\_atoms \gets$ atoms with highest discriminability
    \State $candidates \gets best\_atoms$

    \For{each $atom_1 \in best\_atoms$ not achieving perfect discriminability}
        \For{each $atom_2 \in atoms$ where $atom_2 \neq atom_1$}
            \State $mcc_{and} \gets$ discriminability of $(atom_1 \land atom_2)$
            \State $mcc_{or} \gets$ discriminability of $(atom_1 \lor atom_2)$

            \If{$mcc_{and}$ improves over $atom_1$}
                \State $candidates.\Call{add}{\text{Conjunction}([atom_1, atom_2], mcc_{and})}$
            \EndIf
            \If{$mcc_{or}$ improves over $atom_1$}
                \State $candidates.\Call{add}{\text{Disjunction}([atom_1, atom_2], mcc_{or})}$
            \EndIf
        \EndFor
    \EndFor

    \State \Return candidate with maximum discriminability
\EndFunction
\end{algorithmic}
\end{algorithm}

\Cref{alg:combine-atoms} illustrates the combination process.
For each best atom below perfect discriminability, the algorithm scores its conjunction and its disjunction with every other atom.
To do so, it intersects or unites the sets of vectors in which the two features hold, which directly gives the passing and failing observations that satisfy the combined predicate.
A combination becomes a candidate only if it strictly improves on the atom, and the most discriminative candidate is returned.

\subsubsection{Multi-Candidate Combination}

Several candidate causes may reach the same maximal discriminability after the initial combination of atoms.
\FORECAST{} then attempts to combine them to find an even more discriminative compound cause.
This arises when features correlate with failures independently, but their combination discriminates better.

The multi-candidate combination first identifies all candidates, atoms or already combined predicates, with equal maximal discriminability.
It then tests pairwise combinations of these candidates with both AND and OR, computing the discriminability of the conjunction and of the disjunction of each pair.
A combination is accepted only if it \emph{strictly improves} discriminability beyond the individual candidates.
If no combination improves, the tie-breaking selection described below decides.

The algorithm limits exploration to pairs among the first six candidates.
This avoids exponential complexity while still capturing common multi-feature failure conditions.
Combinations that share features are skipped as redundant, since, for example, combining \texttt{x > 0} with \texttt{(x > 0) AND (y < 5)} would be subsumed.

When already combined candidates are joined, each of them keeps its own operator.
For example, joining \texttt{(A AND B)} with \texttt{(C OR D)} by AND yields \texttt{A AND B AND (C OR D)}, which is both the formula that is scored and the one that is reported.
Joining a conjunction by AND or a disjunction by OR needs no parentheses, as both are associative.

This strategy lets \FORECAST{} find complex failure conditions that require multiple independent features to characterize the bug fully.
An example is ``failure occurs when input is non-numeric AND the operator is subtraction'', which neither feature alone might strongly predict.

\subsubsection{Example: Combining Features}

Returning to our expression parser example (Figure~\ref{fig:expr-parser-bug}), \FORECAST{} extracts atoms at every comparable node.
Consider the entry features of the node of \texttt{parse\_add\_sub()} that \texttt{parse()} calls on Line~23.
Its direct recursive calls also stay at this node (\Cref{sub:traced-relevance-trees}).
No atom separates the runs well there.
The strongest atom, $\phi_1$, states that the condition \texttt{token == "+"} on Line~66 evaluates to \textsc{False}.
The next atoms are weaker still, among them $\phi_2$ for variable \texttt{s}, which is defined on Line~17 and used on Line~19 of \texttt{parse()} before the call.
The algorithm therefore explores combinations:

\begin{align*}
\text{MCC}(\phi_1) &= 0.284 \\
\text{MCC}(\phi_2) &= 0.152 \\
\text{MCC}(\phi_1 \lor \phi_2) &= 0.337
\end{align*}

The disjunction improves over both atoms, so it becomes the entry cause of this node.
It still separates the runs only weakly, however.
Since it stays below the minimum discriminability (\Cref{sub:simplifying}), it does not enter the final cause chain.
Combining atoms thus sharpens the cause found at a node, but it cannot create a separation that the observations at the node do not support.

\subsection{Simplifying Cause Chains}%
\label{sub:simplifying}

The extraction at individual nodes may yield several equally discriminative candidates per node, and the resulting cause chain may contain weak, redundant, or too many causes.
\FORECAST{} therefore simplifies it in five steps, which the following subsections describe in the order in which they are applied.

\subsubsection{Tie-Breaking for Equal Discriminability}

Several candidates at a node may reach the same maximal discriminability without combinations improving it (\Cref{sub:combining}).
\FORECAST{} then reduces them to the single cause most useful for repair agents.
It applies the following tie-breaking criteria lexicographically in this order:

\begin{description}
    \item[Simplicity:] Prefer atoms over conjunctions and disjunctions.
    Simpler causes with fewer constituent features are easier to understand and act upon.
    A single condition like \texttt{value != 0} is preferred over a conjunction like \texttt{(value != 0) AND (Line~69 executed)} if both have equal discriminability.

    \item[Positivity:] Prefer non-negated features.
    Statements about what \emph{is} happening provide more directly actionable information than statements about what \emph{is not} happening.
    For example, \texttt{token == '-'} is preferred over \texttt{NOT (token == '+')} when explaining a subtraction bug.

    \item[Support:] Prefer features that occur more frequently in failing tests.
    Higher support indicates that the feature is consistently present when failures occur, which makes it a more reliable indicator.
    Support is measured as the ratio of failing observations exhibiting the feature to all failing observations.

    \item[Actionability:] Prefer feature types that provide more actionable information based on their semantic content.
    For example, conditions like \texttt{token == '-'} or def-use relationships like \texttt{self.left defined at Line~8 used at Line~19} provide more specific guidance for repair.
    Generic observations like \texttt{Line~42 executed} or symptomatic properties like \texttt{string is ASCII} provide less.
\end{description}

\begin{wraptable}{r}{0.6\textwidth}
    \caption{Feature Actionability Ranking. Higher ranks indicate more actionable information for repair.}%
    \label{tab:actionability}
    \Description{A table that ranks feature types by actionability from 6, the highest, to 1, the lowest. Conditions, def-use pairs, and variable comparisons rank 6. Scalar pairs and return values rank 5. None checks, function errors, lines, and function calls rank 4. Loops and length predicates rank 3, branch decisions rank 2, and string properties rank 1.}
    \begin{center}
    \resizebox{0.58\textwidth}{!}{%
    \begin{tabular}{ll}
    \toprule
    \textbf{Rank} & \textbf{Feature Types} \\
    \midrule
    6 (Highest) & Conditions, Def-Use, Variable comparisons \\
    \rowcolor{tablerow}5 & Scalar pairs, Return values \\
    4 & None checks, Function errors, Lines, Function calls \\
    \rowcolor{tablerow}3 & Loops, Length predicates \\
    2 & Branch decisions \\
    \rowcolor{tablerow}1 (Lowest) & String properties (ASCII, digits, special chars, empty) \\
    \bottomrule
    \end{tabular}
    }
    \end{center}
\end{wraptable}
For causes with several atoms (conjunctions and disjunctions), complexity is measured by the number of constituent atoms, and the negation count sums all negated atoms.
Support is averaged across the atoms, and actionability is their mean rank.
\Cref{tab:actionability} ranks the feature types by actionability, following prior studies~\cite{smytzek2026efdd} showing that certain program properties are more useful for fault localization.
Conditions and def-use relationships rank highest, because they directly describe control flow decisions and data dependencies that often pinpoint bug locations.
String properties rank lowest, as they typically describe symptoms rather than causes.

\subsubsection{Discriminability Threshold}

A cause chain should contain only causes that clearly separate the failing from the passing runs.
\FORECAST{} therefore discards every cause whose discriminability lies below a minimum threshold, by default $0.6$.
Causes below it separate the runs only moderately or weakly and would compete with stronger causes for the limited length of the cause chain (\Cref{sub:chain-length}).
If no cause reaches the threshold, \FORECAST{} keeps all causes, since an empty cause chain would wrongly state that no observed feature separates the runs.
The agent then judges these weaker causes by their discriminability, with a caveat if even the strongest one reaches at most $0.1$ (\Cref{sec:enhancing-repair-agents}).

\subsubsection{Duplicate Elimination}

Causes may appear multiple times in the cause chain, either as exact duplicates or in semantically equivalent forms.
\FORECAST{} computes a unique key for each cause based on its type (Atom, Conjunction, Disjunction) and its constituent features.
It maintains the set of keys seen so far and skips any cause whose key has already been encountered.
Each distinct piece of information thus appears only once in the final cause chain.

\subsubsection{Subsumption Elimination}

\FORECAST{} removes a cause whose features overlap with those of another one, depending on their discriminability.
When causes have \emph{equal discriminability}, the simpler cause is always preferred.
Let cause $D_1$ with features $F_1$ and cause $D_2$ with features $F_2$ have the same discriminability, with $F_1 \subset F_2$.
Then $D_2$ is subsumed by $D_1$ and removed.
For example, if both \texttt{A} and \texttt{A AND B} have discriminability 1.00, \texttt{A} is retained, as it explains as much with less complexity.

The features of one cause may contain those of another while their discriminability differs.
The more discriminative cause is then kept regardless of complexity.
For instance, if \texttt{A} has discriminability 0.75 and \texttt{A AND B} has 1.00, \texttt{A} is removed.
If \texttt{A AND B} has 0.80 and \texttt{B} alone has 1.00, the conjunction is removed.

\subsubsection{Chain Length Management}%
\label{sub:chain-length}

To keep cause chains comprehensible, \FORECAST{} limits their length.
From the causes that remain after the previous steps, it selects the top-$k$ most discriminative ones (by default $k=10$) while preserving their temporal order.
Among equally discriminative causes, divergences are selected first, since the causes observed after such a call describe what follows it.

\begin{algorithm}[t]
\caption{Simplifying Cause Chains}%
\label{alg:simplify}
\begin{algorithmic}[1]
\Function{Simplify}{$candidates, min\_threshold, max\_length$} \Comment{tied best candidates per node}
    \State $causes \gets []$
    \For{each $tied \in candidates$} \Comment{in temporal order}
        \State $causes.\Call{add}{\text{most useful cause in } tied}$ \Comment{simplicity, positivity, support, actionability}
    \EndFor

    \State $filtered \gets$ causes meeting minimum discriminability
    \If{$filtered$ is empty}
        \State $filtered \gets causes$ \Comment{Keep all if none meet threshold}
    \EndIf

    \State $result \gets []$
    \For{each $cause \in filtered$}
        \If{$cause$ is exact duplicate of previous cause}
            \State \textbf{continue}
        \EndIf

        \If{simpler cause with equal discriminability in $result$}
            \State \textbf{continue} \Comment{Subsumed by equal-disc simpler cause}
        \EndIf

        \If{cause with better discriminability subsumes $cause$}
            \State \textbf{continue} \Comment{Subsumed by higher-disc cause}
        \EndIf

        \State remove from $result$ any causes subsumed by $cause$
        \State $result.\Call{add}{cause}$
    \EndFor

    \If{$|result| > max\_length$}
        \State $result \gets$ top-$k$ by discriminability, divergences first on ties (preserve temporal order)
    \EndIf

    \State \Return $result$
\EndFunction
\end{algorithmic}
\end{algorithm}

\Cref{alg:simplify} summarizes these steps.
Because duplicates and subsumed causes are removed before the top-$k$ are selected, they do not occupy its slots, so distinct causes with slightly lower discriminability can be included instead.

\subsubsection{Complete Diagnosis Example}

Returning to the expression parser from Figure~\ref{fig:expr-parser-bug}, we now trace how \FORECAST{} constructs the complete cause-chain diagnosis for its two failing tests.
After extracting features at all comparable nodes and combining atoms, \FORECAST{} obtains an initial cause chain of 19 causes, among them one divergence.
Only four of them reach the minimum discriminability of 0.6.
The other 15 separate the runs at most weakly and are filtered out.
Examples are the disjunction of \Cref{sub:combining} (0.337) and the condition \texttt{tokens and tokens[0] in '*/'} on Line~52 evaluating to \textsc{False} (0.316).
The four remaining causes are distinct, and none subsumes another, so the final cause chain is:

\begin{enumerate}
    \item Condition \texttt{token == "+"} on Line~66 of \texttt{parse.py} evaluates to \textsc{False} (disc: 0.61)
    \item Function \texttt{parse\_add\_sub()} on Line~69 does not return \texttt{None} (disc: 0.61)
    \item The failing run called \texttt{Sub.evaluate()} from \texttt{Mul.evaluate()}, which no passing run called (disc: 0.66)
    \item \texttt{self.left} defined on Line~8 of \texttt{arithmetic.py} is used on Line~24 (disc: 1.00)
\end{enumerate}

This cause chain describes how the failing runs differ from the passing ones.
The first two causes stem from a different node than the disjunction of \Cref{sub:combining}.
At this node, \texttt{parse\_add\_sub()} is called again to parse a parenthesized subexpression.
Of the failing tests, only \texttt{2 * (1 - 1)} reaches it, and of the passing ones, \texttt{1 + (2 * 3)} and \texttt{(1 + 2) * 3}.
In the failing run, the parenthesized expression is a subtraction.
The token on Line~66 is therefore not \texttt{+} (step 1), and the function returns a \texttt{Sub} object from Line~69 (step 2).
The passing runs parse a multiplication or an addition there.
Both causes reach only $0.61$, because every call of \texttt{parse\_add\_sub()} is a separate observation.
The failing run calls it twice at this node, once for \texttt{1 - 1} and once, recursively, for the remaining \texttt{1}, which contains no operator.
The multiplication then evaluates the subtraction, a call that no passing run makes (step 3).
The left operand of this multiplication is the constant \texttt{2}, which it evaluates directly after reading \texttt{self.left} on Line~24.
The left operand of the passing \texttt{(1 + 2) * 3}, in contrast, is a sum (step 4).
The cause chain does not name Line~19 itself.
The passing test \texttt{1 - 0} executes this line, too, and because adding and subtracting zero give the same result, its run takes the same path as the failing ones.
Instead, the cause chain leads from the parsing of the minus sign to the \texttt{Sub} object and to the call of \texttt{Sub.evaluate()} that only a failing run makes, the method that contains the fault.

\section{Enhancing Repair Agents}%
\label{sec:enhancing-repair-agents}

How does \FORECAST{} integrate its cause-chain diagnoses into arbitrary repair agents?
The integration is \emph{agent-agnostic}.
Repair agents differ in how they perform fault localization, plan their actions, or post-process responses.
\FORECAST{} nevertheless injects the cause-chain diagnosis at a single well-defined point: the issue description the agent receives.
From this description, the agent builds every prompt it sends to the underlying LLM.

\subsection{General Repair Agent Pipeline}

\begin{figure}[t]
  \centering
  \begin{tikzpicture}[
    sn/.style={circle, draw, inner sep=0pt, minimum size=3.5mm,
               font=\tiny},
    proc/.style={draw, rounded corners=2pt, fill=black!6,
                 minimum height=9mm, minimum width=2.0cm,
                 align=center, font=\footnotesize},
    diag/.style={proc, fill=forecastblue!18},
    artifact/.style={font=\footnotesize\itshape, align=center},
    arr/.style={-{Stealth[length=2mm]}, thick},
  ]
    \node[diag] (forecast) {\FORECAST{}\\[-1pt]{\scriptsize(Diagnosis)}};
    \node[sn, below=1pt of forecast] (s3) {3};
    \node[artifact, left=0.6cm of forecast] (tests) {Program\\[-2pt]and Tests};
    \node[sn, below=1pt of tests] (s2) {2};

    \node[circle, draw, thick, inner sep=0pt, minimum size=3mm,
          font=\footnotesize, above=0.9cm of forecast] (plus) {$+$};

    \node[artifact] (issue) at (tests |- plus) {Issue};
    \node[sn, above=1pt of issue] (s1) {1};

    \node[proc, right=0.7cm of plus] (pre) {Pre-\\[-1pt]computation};
    \node[sn, above=1pt of pre] (s4) {4};
    \node[proc, right=0.6cm of pre] (prompt) {Prompt\\[-1pt]Construction};
    \node[sn, above=1pt of prompt] (s5) {5};
    \node[proc, right=0.6cm of prompt] (llm) {LLM\\[-1pt]Query};
    \node[sn, above=1pt of llm] (s6) {6};
    \node[proc, right=0.6cm of llm] (post) {Post-\\[-1pt]processing};
    \node[sn, above=1pt of post] (s7) {7};

    \node[artifact, right=0.7cm of post] (patch) {Patch};
    \node[sn, above=1pt of patch] (s8) {8};

    \draw[arr] (issue) -- (plus);
    \draw[arr] (plus) -- (pre);
    \draw[arr] (pre) -- (prompt);
    \draw[arr] (prompt) -- (llm);
    \draw[arr] (llm) -- (post);
    \draw[arr] (post) -- (patch);
    \draw[arr] (tests) -- (forecast);
    \draw[arr] (forecast) -- node[left, align=right, font=\scriptsize\itshape] {appends\\[-2pt]diagnosis} (plus);

    \node[draw, thick, rounded corners=4pt, inner xsep=6pt, inner ysep=5pt,
          fit=(pre) (post) (s4)] (agent) {};
    \node[font=\footnotesize\bfseries, anchor=south west]
      at ([yshift=1pt]agent.north west) {Repair Agent};
  \end{tikzpicture}
  \caption{Abstract repair agent pipeline.
    Given an issue~\circnum{1}, a repair agent performs pre-computation such as fault localization~\circnum{4},
    constructs a prompt~\circnum{5}, queries the LLM~\circnum{6},
    and post-processes its response~\circnum{7} into a patch~\circnum{8}.
    \FORECAST{}~\circnum{3} computes a diagnosis from the program and its tests~\circnum{2} and appends it to the issue,
    so the diagnosis reaches every prompt the agent constructs from it.}%
  \label{fig:agent-pipeline}
  \Description{A flow diagram. An issue (1) passes a plus sign into a box labeled repair agent with four stages, pre-computation (4), prompt construction (5), LLM query (6), and post-processing (7), which produces a patch (8). Below the issue, FORECAST (3) computes a diagnosis from the program and its tests (2) and appends it to the issue at the plus sign.}
\end{figure}
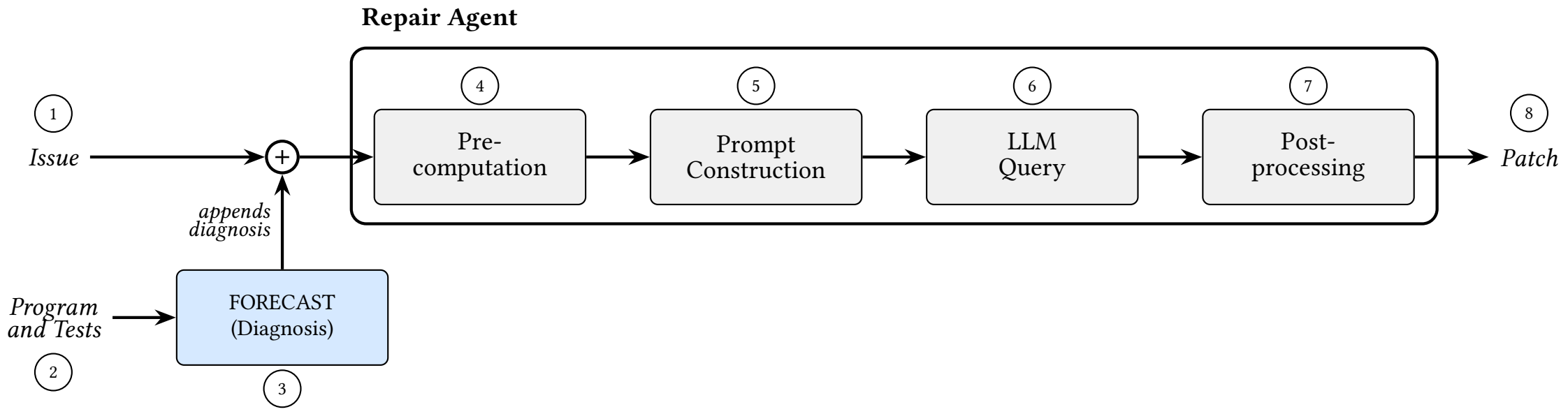

Despite their architectural differences, modern LLM-based repair agents share a common abstract pipeline.
Figure~\ref{fig:agent-pipeline} illustrates this pipeline and shows where \FORECAST{} intervenes.

The pipeline consists of four stages.
In \emph{pre-computation}, before or while constructing its prompts, an agent gathers information about the failure.
This may include fault localization, which identifies the files and lines likely responsible for the failure.
It may also include the retrieval of relevant code context or building an understanding of the project structure, with or without the LLM.
For example, \AUTOCODEROVER{} and \REPAIRAGENT{} receive a spectrum-based fault localization, \AGENTLESS{} narrows the repository down to files and functions, and \SWEAGENT{} issues exploratory commands through its agent-computer interface.
In \emph{prompt construction}, the agent assembles this information into a textual prompt.
The prompt typically contains the problem description, relevant code snippets, and instructions for the repair task.
In the \emph{LLM query}, the prompt is sent to the LLM, which responds with one or more suggested fixes.
In \emph{post-processing}, the agent extracts the patch, may validate it against tests, and may rank or filter candidates.
Moreover, these steps may be repeated, based on the agent's design.

\subsection{Diagnosis Injection}

\FORECAST{} intercepts the pipeline at its input.
Before the agent starts, \FORECAST{} appends a structured \emph{Failure Diagnosis} block to the issue description.
This block contains the cause chain inferred by the algorithm of \Cref{sec:debugging}, serialized as plain text.
Since an agent builds its prompts from the issue, the block appears exactly once in every prompt that presents the issue to the LLM.
The injection is transparent to the agent, and none of its stages requires any modification.
The agent operates exactly as it would without \FORECAST{} but receives a richer issue description.
A cause-chain diagnosis, however, does not always contain causes, and the ways it can fall short are not equivalent.

The inference may run without finding any feature that separates the failing execution from the passing ones.
The resulting cause chain is then empty.
We report that emptiness rather than suppress it.
An empty cause chain states that the executions are indistinguishable over everything \FORECAST{} observes.
Consequently, the failing run neither reaches different code nor takes a different branch than the runs that pass.
That is a constraint on where the fault can be, so the injected block says so.
It directs the agent towards what such a result leaves open.
This includes values entering from outside the traced code, behavior that is \emph{absent} rather than wrong, and results that differ although the same code ran.
From the agent's side, leaving the prompt unchanged would be indistinguishable from never having consulted \FORECAST{} at all.

A cause chain may also exist while each of its causes separates the executions only marginally, with a discriminability of at most $0.1$.
The block then reports the cause chain together with its highest discriminability.
It asks the agent to treat the cause chain as a starting point rather than as the identified cause.
Presenting a weak signal in the same terms as a strong one would invite the agent to chase noise.
The two thresholds serve different purposes.
The minimum discriminability of $0.6$ (\Cref{sub:simplifying}) decides which causes a cause chain keeps when strong ones exist.
If none exists, the strongest available causes are kept rather than leaving the agent without a diagnosis.
The caveat marks only cause chains whose strongest cause is nearly uncorrelated with the failure.
Causes between $0.1$ and $0.6$ still separate the runs moderately and are shown with their discriminability, from which the agent can judge their strength.

The issue is left unchanged only when no cause-chain diagnosis is produced at all.
This happens when collection itself fails, so that we observe nothing about how the executions compare.
The agent then degrades to its baseline behavior.
Reporting indistinguishable executions in that case would assert a finding we never made.

\subsection{Prompt Structure}

To make the augmentation concrete, consider the issue an agent receives for the expression parser of \Cref{fig:expr-parser-bug}.
In \FORECAST{} mode, the problem statement is followed by a \textit{Failure Diagnosis} block in Markdown:

\begin{quote}
\small
\texttt{[problem statement]}

\smallskip\noindent
\texttt{\#\# Failure Diagnosis}

\smallskip\noindent
The following diagnosis shows the cause-effect chain leading to the test failure:

\smallskip\noindent
1.~Condition \texttt{"token == '+'"} at \texttt{expr/parse.py:66} evaluates to False (discriminability: 0.61)\\
\phantom{1.~}LEADS TO\\
2.~Function \texttt{parse\_add\_sub} in \texttt{expr/parse.py:69} does not return None (discriminability: 0.61)\\
\phantom{1.~}LEADS TO\\
3.~The failing run called \texttt{expr/arithmetic.py:Sub.evaluate} from \texttt{expr/arithmetic.py:Mul.evaluate}, which no passing run called (discriminability: 0.66)\\
\phantom{1.~}LEADS TO\\
4.~Variable \texttt{"self.left"} defined at \texttt{expr/arithmetic.py:8} is used at \texttt{expr/arithmetic.py:24} (discriminability: 1.00)
\end{quote}

\noindent Every cause names the file and line where it was observed and carries its discriminability.
Divergences (\Cref{sub:features}) name the calling and the called function.
Apart from the notes for empty and weak cause chains, the block ends with the cause chain itself and adds no instruction on how to use it.
The instructions the agent gives its model thus remain the only ones the model receives.
Wherever the agent places the issue in a prompt, be it a system message, a task description, or a retrieval query, the cause-chain diagnosis follows it there.

\subsection{Generality}

\FORECAST{} treats the issue description as an opaque string and appends the diagnosis block to it, leaving the agent's own prompts untouched.
The approach thus applies to any agent that (1) takes a textual issue description and (2) builds its LLM prompts from it.
It requires no assumptions about the agent's internal structure, the specific LLM, or the programming language of the subject program.
Different agents may nevertheless benefit differently from the injected information, which \Cref{sec:evaluation} measures for five agents of different designs.

\section{Implementation}%
\label{sec:implementation}

\FORECAST{} is implemented as an open-source Python package.
Its modules mirror the pipeline stages of \Cref{sec:debugging} and \Cref{sec:enhancing-repair-agents}, namely instrumentation, trace collection, tree construction, diagnosis inference, and agent integration.
Repair agents, model backends, fault localization formulas and techniques, and subject types are all registered through a plugin interface, so adding a new one requires no change to the core pipeline.

\paragraph{Instrumentation and Trace Collection.}
Instrumentation builds on \SFLKIT{}~\cite{smytzek2022sflkit}, which rewrites the source of the program under test so that it reports whatever we want to observe.
Such events include a function being entered or left, a branch being taken, a condition being evaluated, or a variable being defined or used.
The rewriting only inserts calls that report these events.
Apart from its runtime cost, it thus leaves the program's behavior unchanged, except for programs that inspect their own source or depend on timing.
Every test runs in isolation and produces its own trace, labeled with the test's outcome so that tests can be traced in parallel without any coordination between them.
The traces are then grouped by outcome and handed to the tree builder.

\paragraph{Keeping Tracing Affordable.}
Tracing a program is far more expensive than merely running it, because a trace records every single observation rather than a summary.
Left unchecked, individual tests in our subjects produced traces of hundreds of gigabytes and exhausted their time budget before finishing.

The dominant cost turned out not to be the tests themselves.
Every test is executed in a fresh, isolated environment.
The first test to run in such an environment pays for whatever one-time initialization the project performs.
One of our subjects, for instance, builds a cache of the fonts installed on the system, scanning and parsing each of them.
That work is negligible when it happens once.
Since each test starts from a clean environment, however, it is repeated for every test, and under instrumentation, it is recorded in full.
On one subject, the same test took 66 seconds and produced 1.1\,GB in a fresh environment, against 4 seconds and 54\,MB once the initialization had already happened.
\FORECAST{} therefore performs this initialization once, while preparing the subject's environment and before instrumenting it.
The cost is thus paid a single time and never observed.

Beyond that, \FORECAST{} bounds what a single run may record, since its analysis never needs the full detail.
Predicates over values ask coarse questions, such as whether a string is empty or contains a digit.
Recording a bounded prefix of a value (by default 1\,KB) is therefore enough, and lengths are observed separately and remain exact.
Loop features distinguish only whether a loop ran never, once, or more than once, so observing more than two iterations of a loop adds nothing at all.
These bounds matter for programs that move large values or run exhaustive and nested loops, and they are inexpensive insurance elsewhere.
A configurable size budget (by default 2\,GB per trace before compression) additionally cuts a trace off once it grows too large.
Because observations are recorded whole, a truncated trace remains readable and describes a prefix of the run.
Finally, traces are compressed with gzip as they are written, which is effective because they are highly repetitive.

The second lever is to trace fewer tests.
A cause-chain diagnosis only compares executions at functions that both a passing and a failing run reach (\Cref{sub:traced-relevance-trees}).
A passing test that shares no executed code with the failing tests therefore cannot influence the result and would only cost a traced run.
\FORECAST{} identifies those tests from plain coverage, which is cheap to collect, and traces only the passing tests whose execution overlaps the failing ones.
It prefers the tests that cover most of the failing code.
Among equally overlapping tests, it prefers those that execute the least code besides it, since these give the sharpest contrast and are also the cheapest to trace.
An optional cap keeps only the highest-ranked ones.
Measuring coverage costs a run of its own, so for few candidate tests, \FORECAST{} skips the selection and traces them all.
The selection never empties the set of passing tests.
Should no candidate overlap the failing code, or should coverage fail, a bounded prefix of the candidates is traced instead.
A cause-chain diagnosis without passing runs would have nothing to contrast against.
For the same reason, when an instance lists no passing test at all, \FORECAST{} runs the repository's tests closest to the failing ones and uses those that pass.

\paragraph{Diagnosis Inference.}
Tree construction merges all traces into a single call tree, following function entries and exits while accumulating the observations made between them.
The failing traces span the tree.
Passing traces are then folded in only along the call paths the failing runs take, since a node that no failing run reaches cannot be comparable.
They can be processed in parallel and merged.
The diagnosis stage then realizes the algorithms of \Cref{sec:debugging}.
It marks comparable nodes bottom-up by testing whether passing and failing runs share any observation at a node.
It extracts cause chains depth-first, processing entry observations, then nested calls, then exit observations, so that the cause chain follows the order of execution.
Discriminability is scored with the Matthews correlation coefficient, which is undefined when a feature or its complement is constant across all runs.
Such features carry no information and are scored as uncorrelated.
Feature combinations are evaluated by intersecting (\emph{and}) or uniting (\emph{or}) the sets of observations that satisfy each atom.
To keep the search tractable, every pair contains one of the strongest atoms, and tied candidates are combined only among the first six.
The resulting cause-chain diagnosis is serialized to JSON, Markdown, or plain text.

\paragraph{Agent Integration.}
The integration layer appends the serialized cause chain to the issue description an agent receives (\Cref{sec:enhancing-repair-agents}) and leaves the agent itself unmodified.
Because the cause-chain diagnosis enters at that single point, an agent's own fault localization, planning, and patch generation stay untouched.
An agent that receives no diagnosis behaves exactly as its baseline.
Every prompt in which an agent presents the issue to the model therefore carries the cause chain exactly once.
Agents that query the model many times, such as \SWEAGENT{} and \AUTOCODEROVER{}, thus stay guided throughout their repair loop.
All model calls of all agents pass through one \FORECAST{} model interface, which records every request and its token usage uniformly.
Every patch is judged by the official \SWEBENCH{} harness, never by the agent.

\section{Evaluation}%
\label{sec:evaluation}

We evaluate how much \FORECAST{} improves repair agents, where the improvement comes from, and what it costs.
To this end, we address the following research questions:

\begin{questions}[topsep=2pt]
    \item\label[question]{rq1} \textbf{Effectiveness}. To what extent does the cause-chain diagnosis of \FORECAST{} improve the repair success of existing repair agents, and does the improvement persist across repeated runs?
    \item\label[question]{rq2} \textbf{Diagnosis Components}. Which parts of the cause-chain diagnosis matter, and how does it compare to other debugging aids?
    \item\label[question]{rq3} \textbf{Fix Quality}. How does the cause-chain diagnosis affect the minimality, applicability, and correctness of the generated fixes?
    \item\label[question]{rq4} \textbf{Efficiency}. What does computing a cause-chain diagnosis cost, and how does the diagnosis change the tokens, cost, and time of the repair?
\end{questions}

\subsection{Experimental Setup}%
\label{sec:setup}

\paragraph{Study Design}
To answer our research questions, we conduct a controlled, paired experiment.
We run each repair agent in two modes whose inputs differ in exactly one factor: the presence of a cause-chain diagnosis.
In the \emph{baseline} mode, the agent receives only the inputs it was originally designed for, namely the issue description and access to the repository under test.
In the \emph{\FORECAST{}} mode, the agent receives the identical inputs plus the cause-chain diagnosis (\Cref{sec:debugging}), injected into its prompt (\Cref{sec:enhancing-repair-agents}).
The agent, the benchmark, the underlying model, and all decoding parameters stay fixed between the two modes.
This isolates the effect of the diagnosis on repair effectiveness (\Cref{rq1}), fix quality (\Cref{rq3}), and efficiency (\Cref{rq4}).
Two further experiments examine this effect.
We repeat the runs of both modes to measure how much their outcomes vary from run to run and whether the benefit persists in each repetition (\Cref{rq1}).
We also replace the diagnosis with variants that withhold or substitute parts of it (\Cref{rq2}).

\paragraph{Benchmark}
We evaluate on \SWELITE{}~\cite{jimenez2024swebench}, a curated subset of $\nInstances$ real-world GitHub issues drawn from $\nProjects$ popular open-source Python projects.
Each instance provides an issue description (\emph{problem statement}), the repository pinned at the buggy \emph{base commit}, and a hidden developer \emph{test patch} that adds the tests exercising the fix.
It also provides two sets of test identifiers.
\texttt{FAIL\_TO\_PASS} lists the tests that must change from failing to passing for the issue to be resolved.
\texttt{PASS\_TO\_PASS} lists the tests that must keep passing to guard against regressions.
We use \SWELITE{} because it is the de facto standard for evaluating LLM-based repair agents and remains representative of real maintenance tasks.
It is also small enough to run every agent in both modes within our compute budget.
We detail the resulting subject characteristics in \Cref{sec:subjects}.

\paragraph{Repair Agents}
We study five widely used, structurally diverse repair agents.
\AUTOCODEROVER{}~\cite{zhang2024autocoderover} combines program-structure-aware code search with spectrum-based fault localization.
\AGENTLESS{}~\cite{xia2025agentless} follows a fixed localize--repair--validate pipeline without an autonomous agent loop and validates its patches with reproduction tests it generates from the issue.
\SWEAGENT{}~\cite{yang2024sweagent} drives an LLM through an agent-computer interface to navigate, edit, and test the repository interactively.
\MINISWEAGENT{}~\cite{sweagent2025minisweagent} is a minimal autonomous agent by the authors of \SWEAGENT{} that interacts with the repository through shell commands alone.
\REPAIRAGENT{}~\cite{bouzenia2025repairagent} drives its repair through hypotheses about the bug and runs of the developer's tests.
We include \AGENTLESS{} and \REPAIRAGENT{} specifically because they already work with tests in their baseline mode, which lets us control for test access (see below).
These agents span the spectrum from agentless pipelines to fully autonomous agents.
Improvements that hold across all of them are thus unlikely to be artifacts of a single architecture.

We run \SWEAGENT{}, \MINISWEAGENT{}, and \AGENTLESS{} unmodified, from the authors' published code at a pinned version.
\SWEAGENT{} and \MINISWEAGENT{} use the configurations their authors use for \SWEBENCH{}.
\AGENTLESS{} replays its published \SWEBENCH{} pipeline stage by stage, with localization, 40 sampled patches, regression and generated reproduction tests, and reranking.
\AGENTLESS{} deviates in two respects.
Its embedding-based file retrieval relies on an embedding model that our models do not offer.
File localization is therefore left to the language model alone, a configuration \AGENTLESS{} supports.
In addition, each of its steps may answer in up to $4{,}096$ tokens rather than its original $300$ to $1{,}024$.
With the original limits, a model that reasons before it answers is cut off.
Localization then names no file, and repair and test samples end inside their edit or test.
Both modes run with the raised limits.

\AUTOCODEROVER{}'s code is not openly licensed, so we completely re-implement it from its paper.
The re-implementation comprises stratified context retrieval over the paper's seven search APIs and method-level spectrum-based fault localization, which the agent sees before retrieval.
It also comprises patch generation with a retry loop that checks format, applicability, syntax, and the tests.
\FORECAST{} computes fault localization itself as an Ochiai ranking~\cite{abreu2006ochiai} over the very test executions it traces, and provides its five top-ranked methods in both modes.
\AUTOCODEROVER{}'s baseline thus already receives a test-based localization signal derived from the same data as the cause-chain diagnosis.
The tests its retry loop runs are the instance's \texttt{FAIL\_TO\_PASS} and \texttt{PASS\_TO\_PASS} tests with the developer's test patch applied, again in both modes.
Like \REPAIRAGENT{} below, \AUTOCODEROVER{} thus has access to the tests from which the diagnosis is derived.

Each agent runs inside the instance's \SWEBENCH{} Docker image.
\SWEAGENT{}, \MINISWEAGENT{}, and \AGENTLESS{} receive only the issue, as in their original setting, and the developer's tests stay hidden from them.
The inputs of the baseline and \FORECAST{} modes of an agent differ in exactly one factor: the injected cause-chain diagnosis.
\FORECAST{} computes it from the tests before the repair starts.

\paragraph{Controlling for Test-Suite Access}
\FORECAST{} derives its cause-chain diagnosis from the passing and failing test executions.
Any improvement could therefore merely reflect that the agent receives additional information extracted from tests, rather than the value of the cause-chain diagnosis itself.
To guard against this confound, we include two agents whose baselines already exploit tests, one for each of the two settings in which an agent may work.
\AGENTLESS{} represents the setting in which an issue comes without a failing test.
In both modes, it generates reproduction tests from the issue and selects one that reproduces the issue by majority vote.
To choose among its candidate patches, it runs this test together with the repository's existing tests that pass before any change.
Its baseline thus already has a failing test derived from the issue, as a developer or a test generator would write it.
\REPAIRAGENT{}~\cite{bouzenia2025repairagent} represents the setting in which the developer's tests are available.
It is an autonomous, test-driven repair agent that runs and inspects the test suite as part of its repair loop.
In its baseline mode, it thus already has full, unrestricted access to the same tests from which \FORECAST{} builds its diagnosis.
\AUTOCODEROVER{}, which validates its patches against these tests as well, complements it.
If \FORECAST{} still improves these agents, the gain cannot be attributed to mere access to tests, since their baselines already exploit them.
It instead reflects what the cause-chain diagnosis adds to the test outcomes, namely how the failing execution differs from the passing ones.

\REPAIRAGENT{} was originally proposed and evaluated on the \DEFECTS4J{}~\cite{just2014defects4j} Java benchmark, and its tools are \DEFECTS4J{} commands for Java projects.
We therefore port it to Python.
We keep its design, including the dynamic prompt, the three-state machine with hypothesis tracking, its tool set, its middleware, and its budget of 40 commands.
We also keep its prompts, adapted from Java to Python.
We re-target its tools to the \SWEBENCH{} image, where they run the \texttt{FAIL\_TO\_PASS} and \texttt{PASS\_TO\_PASS} tests with the developer's test patch applied.
By default, \REPAIRAGENT{} receives \emph{perfect} fault localization: the lines the developer changed.
On \SWELITE{}, this would reveal the fix.
We therefore use its alternative of an existing fault localization tool, the same spectrum-based ranking that \AUTOCODEROVER{} receives, and show \REPAIRAGENT{} the ten top-ranked methods.
We run it in the same two modes and with the same models as the other agents.

\paragraph{Underlying Models}
We run every agent, in both modes, with each of two large language models.
\ANTHROPIC{}'s \HAIKU{} (\texttt{claude-haiku-4-5-20251001})~\cite{anthropic2025haiku45} is a commercial model that we access through its API.
\QWEN{}~27B (\texttt{qwen3.5:27b}) is an open-weight model that we serve locally with Ollama.
Each agent keeps the decoding settings of its published configuration, so that it behaves as its authors evaluated it.
\AGENTLESS{}, for instance, samples candidate patches at temperature $0.8$.
Some configurations set both a temperature and a nucleus-sampling threshold.
\HAIKU{} then receives only the temperature, as its API accepts only one of the two.
\QWEN{} runs with a context window of $131{,}072$ tokens and with its thinking mode disabled, as \HAIKU{} does not think unless asked to.
The two modes of an agent share the model and its settings, so the measured difference stems from the injected cause-chain diagnosis rather than from model or configuration differences.
Every LLM call of every agent passes through \FORECAST{}'s model interface, which records its tokens.
We price the tokens of \HAIKU{} at \ANTHROPIC{}'s list prices, independent of any discounts.
These are \$1 per million input and \$5 per million output tokens, and \$1.25 and \$0.10 per million tokens written to and read from the prompt cache.
\QWEN{}, which we serve locally, incurs no charge per token, although its hardware and energy are not free, so we compare its runs by their tokens.
Two agents have a budget.
\SWEAGENT{} and \MINISWEAGENT{} stop a run once it has cost \$3, or after $150$ model calls and $250$ steps, respectively.
\SWEAGENT{} then submits the changes made so far, whereas \MINISWEAGENT{} submits nothing.
For \QWEN{}, the agents count its tokens at the same prices, without prompt caching, which Ollama does not provide.
Its budget thus amounts to about three million tokens per run.
Runs of \HAIKU{}, whose prompts are mostly read from the cache, can use several times as many.

\paragraph{Diagnosis and Execution Environment}
For every instance, \FORECAST{} first derives a cause-chain diagnosis from the program's passing and failing executions.
We run each \texttt{FAIL\_TO\_PASS} and \texttt{PASS\_TO\_PASS} test inside the official \SWEBENCH{} per-instance Docker image, which fixes the operating system, interpreter, and dependency versions.
The traces are thus collected in the same environment in which patches are later validated.
The developer test patch is applied before tracing so that the failing tests are present and exercise the buggy code.
We instrument the application code with \textsc{SFLKit}~\cite{smytzek2022sflkit} to record fine-grained execution features~\cite{smytzek2026efdd}.
Test files, configuration files, and test helpers are explicitly excluded from instrumentation, so the resulting diagnoses point at the code under repair rather than at the tests.
From the collected event traces, \FORECAST{} builds a traced relevance tree and derives the cause-chain diagnosis (\Cref{sec:debugging}).
The diagnosis is cached and reused across both the prompt-generation and the repair stages.
Tests are executed with a $120$\,s per-test timeout and up to four parallel workers.

Some instances list several hundred \texttt{PASS\_TO\_PASS} tests, and tracing all of them is not affordable.
We therefore apply the trace bounds and the coverage-based test selection of \Cref{sec:implementation} throughout, in \emph{both} modes and for every subject, so that the comparison remains controlled.
Instances with at most $25$ candidate passing tests are traced in full (which is true for $101$ instances of \SWELITE{}).
For the others, coverage is measured for at most $200$ candidates, those in the failing tests' files first, and the $120$ top-ranked overlapping tests are traced ($166$ instances).
Where coverage of the failing tests could not be measured, all measured candidates are traced ($31$ instances), and where coverage failed altogether, the first $120$ are traced ($2$ instances).
For the six instances that list no \texttt{PASS\_TO\_PASS} test, \FORECAST{} recovers passing tests from the repository as described in \Cref{sec:implementation}.
In total, we trace $\nFailingTraced$ failing and $\nPassingTraced$ passing test executions, a median of $1$ failing and $42.5$ passing tests per instance.
The selection uses only the tests, including those of the developer's test patch, and never the developer's fix.
It thus gives the diagnosis no information beyond the tests themselves.
With these bounds, \FORECAST{} produced a cause-chain diagnosis for every one of the $\nInstances$ instances, and no diagnosis relies on the developer's fix.

\paragraph{Repair Pipeline}
Each repair attempt proceeds through a fixed pipeline of six stages.
\begin{enumerate*}[label=(\arabic*)]
    \item \FORECAST{} first \emph{collects} execution events for the failing and passing tests,
    \item \emph{builds} the traced relevance tree, and
    \item \emph{diagnoses} to obtain the cause chain.
    \item The \emph{prompt} stage assembles the agent's native prompt and, in \FORECAST{} mode, injects the cause-chain diagnosis.
    \item The \emph{run} stage is a single LLM-driven repair attempt that produces a unified-diff patch (pass@1).
    \item The \emph{evaluate} stage applies the patch inside the instance's Docker image and executes the full \texttt{FAIL\_TO\_PASS} and \texttt{PASS\_TO\_PASS} suites.
\end{enumerate*}
Stages (1)--(3) run once per instance, before any repair, and all agents and both models receive the same diagnosis.
Only the \FORECAST{} mode receives it.
The fault localization of \AUTOCODEROVER{} and \REPAIRAGENT{}, computed from the same traces, is provided in both modes.
Each repair attempt is limited to six hours.
An attempt that yields no verdict counts as unresolved, as does one in which the agent fails.
In the main experiment, each agent, mode, and model is run once per instance, resulting in a total of $\nAttempts$ repair attempts.
The repeated runs and the variants of the diagnosis come on top of these and are described below.

\paragraph{Repeated Runs}
The agents do not behave deterministically.
Some of them sample at a positive temperature, and even greedy decoding can vary with batching on the serving side.
A second run of the same configuration may therefore resolve different instances.
To measure this variation, we run every agent with \QWEN{} twice more in each mode, with the same cause-chain diagnoses and settings and on all $\nInstances$ instances.
Together with the main experiment, each agent and mode thus has three runs with \QWEN{}.
With \HAIKU{}, whose tokens we pay for, we repeat only the \FORECAST{} mode, once per agent.
This shows how much the \FORECAST{} mode varies with the commercial model.
The runs of a mode are interchangeable, so we measure the gain of the diagnosis between every \FORECAST{} run and every baseline run of an agent.

\begin{table}[t]
    \caption{Variants of the diagnosis.
    Each variant replaces the cause chain in the \textit{Failure Diagnosis} block and leaves everything else as in the \FORECAST{} mode.
    The recomputed variants run only on the instances whose chain they change.}%
    \label{tab:variants}
    \Description{A table that lists the thirteen variants of the diagnosis, what the agent receives instead of the cause chain in each, and whether the variant ran with both agents or only with mini-SWE-agent.}
    \small
    \begin{tabular}{@{}lp{0.62\textwidth}l@{}}
    \toprule
    \textbf{Variant} & \textbf{What the agent receives instead of the chain} & \textbf{Agents} \\
    \midrule
    Foreign               & the chain of another instance of the same project & both \\
    \rowcolor{tablerow}Unordered             & the same causes in a random order, as a list without \texttt{LEADS TO} & both \\
    Locations             & only the file, line, or function each cause names, with its discriminability & both \\
    \rowcolor{tablerow}No divergences        & the chain recomputed without divergences ($147$ instances) & both \\
    Atoms                 & the chain recomputed without conjunctions and disjunctions ($125$ instances) & both \\
    \rowcolor{tablerow}Test output           & the failing tests and their output on the unchanged code & \MINISWEAGENT{} \\
    SBFL                  & the ten methods spectrum-based fault localization ranks highest & \MINISWEAGENT{} \\
    \rowcolor{tablerow}Chain and test output & the chain, followed by the failing tests and their output & \MINISWEAGENT{} \\
    Wording               & the chain, introduced as observations and joined by \texttt{THEN} & both \\
    \rowcolor{tablerow}$k = 5$, $k = 20$     & the chain recomputed with at most $5$ or $20$ instead of $10$ causes & both \\
    $t = 0.4$, $t = 0.8$  & the chain recomputed with a minimal discriminability of $0.4$ or $0.8$ instead of $0.6$ & both \\
    \bottomrule
    \end{tabular}
\end{table}

\paragraph{Variants of the Diagnosis}
To find out which parts of the cause-chain diagnosis the agents benefit from, we replace the cause chain in the \textit{Failure Diagnosis} block with variants (\Cref{tab:variants}).
Each variant withholds or substitutes some of the information of the cause chain.
Everything else stays as in the \FORECAST{} mode.
A variant that injects something other than a cause chain introduces it with a sentence of its own, for instance ``The following tests fail on the current code, with their output:''.

Three variants test whether the content of the cause chain matters.
\emph{Foreign} injects the cause chain of another instance of the same project.
It is a negative control that is as structured and plausible as a real diagnosis but says nothing about the failure at hand.
The donor of an instance is a single fixed one; it is the instance half the list further on in the sorted list of the project's instances, wrapping around at the end.
Every instance thus receives the cause chain of another instance of its own project.
\emph{Unordered} presents the same causes in a random order, as a list without \texttt{LEADS TO}, and thus tests whether their order of execution matters.
Each instance gets one random permutation, seeded by the instance, which always differs from the order of execution, and the causes appear as an unnumbered list.
\emph{Locations} keeps only the file, line, or function each cause names, with its discriminability.
It thus tests whether the conditions, values, and data flows of the causes add to where they were observed.

Two variants recompute the cause chain from the traced relevance tree without one kind of cause.
\emph{No divergences} omits the calls that only failing runs make.
\emph{Atoms} omits conjunctions and disjunctions, so that each cause is a single feature.
Only the $147$ and $125$ instances whose cause chain contains such a cause are affected.
For the others, the prompt is the same as in the \FORECAST{} mode, and we take its outcome.

Three variants compare the cause chain with information that is available without \FORECAST{}.
\emph{Test output} injects the failing tests and their output, such as tracebacks and assertion messages.
This output is recorded on the unchanged code with the developer's test patch applied and cut to about $4{,}000$ characters, about the size of a long cause chain.
\emph{SBFL} injects the ten methods that spectrum-based fault localization over the same executions ranks highest, the ranking that \AUTOCODEROVER{} and \REPAIRAGENT{} receive in both modes.
\emph{Chain and test output} injects the cause chain followed by the test output, to find out whether the two complement each other.

\emph{Wording} keeps the cause chain but introduces it as observations that distinguish the failing from the passing runs, in the order in which they occur.
It also joins the causes by \texttt{THEN} instead of \texttt{LEADS TO}.
It thus tests whether the agents benefit from the presentation of the causes as a cause-effect chain (\Cref{sec:construct-validity}).
Finally, we vary the two parameters of the simplification (\Cref{sec:debugging}).
These are the maximal number of causes ($k = 5$ and $k = 20$ instead of $10$) and the minimal discriminability ($t = 0.4$ and $t = 0.8$ instead of $0.6$).
Changing the parameters may lead \FORECAST{} to consider different features.
A recomputed cause chain can therefore differ from the evaluated one in more than its length, and the same holds for the cause chains without divergences or combinations.
As for these, only the instances whose cause chain changes are run again ($283$, $282$, $290$, and $288$ instances for $k = 5$, $k = 20$, $t = 0.4$, and $t = 0.8$, respectively).

We run the variants with \QWEN{}, which we serve locally, and with two agents.
\MINISWEAGENT{} interacts with the repository through shell commands alone and never sees the developer's tests.
The injected block is thus the only information it receives from them, which makes it a direct probe of what the block conveys.
\REPAIRAGENT{} runs the developer's tests and receives the spectrum-based ranking in both modes, so it shows what a variant adds to the information an agent already has.
For the same reason, \REPAIRAGENT{} does not run the test output, SBFL, and chain and test output variants, whose added information its baseline already receives.

\paragraph{Metrics}
Across all research questions, we compare the baseline and \FORECAST{} mode of each agent on identical instances.
We report both the absolute and the relative change, so that improvements are interpretable independently of an agent's starting point.
Our primary outcome is whether an instance is \emph{resolved}.
For that, the agent's patch must apply cleanly to the repository, turn every \texttt{FAIL\_TO\_PASS} test from failing to passing, and leave every \texttt{PASS\_TO\_PASS} test passing.
This is the official \SWEBENCH{} resolution criterion.
We evaluate every patch by applying it inside the instance's Docker image and running both test sets there.
Resolution is thus judged in the same environment that we use to collect the cause-chain diagnosis.

For \Cref{rq1}, we report the number and percentage of resolved instances, both per agent and aggregated across agents.
We contrast each agent in \FORECAST{} mode with its baseline to isolate the benefit of the diagnosis.
For the change in percentage points, we give a $95\%$ confidence interval from a paired bootstrap over the instances with $10{,}000$ resamples.
Both modes of an agent run on the same instances.
We therefore test whether their resolution rates differ with the exact two-sided McNemar test~\cite{mcnemar1947} on the paired per-instance outcomes.
Over all agents of a model, the outcomes for the same instance are not independent.
We therefore test the totals with the exact two-sided sign test over the instances, which counts the instances that more agents resolve with the diagnosis and those that fewer agents resolve with it.
With twelve comparisons (five agents plus their combined resolve-rate and two models), we apply the Bonferroni correction and multiply each $p$-value by the number of comparisons.
We consider a difference significant if the corrected $p$-value is below $0.05$.
In the tables, we underline the differences that are significant.

To assess fix quality (\Cref{rq3}), we characterize the generated patches by their \emph{minimality} and their \emph{applicability}.
We measure minimality as the number of changed lines (added plus removed) and the number of distinct files a patch modifies.
Applicability is the fraction of generated patches that apply cleanly to the repository without error.
Smaller, cleanly applicable patches are easier to review and less likely to carry unrelated changes.
Agents that work in the repository sometimes leave files behind that they created for themselves, such as reproduction scripts.
We count those new files separately and measure minimality over the changes to files that exist in the repository.

For efficiency (\Cref{rq4}), we weigh the computational overhead of producing a cause-chain diagnosis against the repair cost.
We measure the overhead as the wall-clock time of the event-collection, tree-building, and diagnosis stages.
We measure the repair cost as the LLM tokens consumed and the end-to-end repair time per instance.
For all of these measures, we report the distribution across instances rather than only aggregates.
The effect of the diagnosis on outliers, such as unusually large patches or expensive diagnoses, thus remains visible.
We additionally record the \emph{discriminability} of each diagnosis.
This observation-level score quantifies how sharply its cause chain separates the observed passing from the observed failing executions (\Cref{sec:debugging}).
We use it to characterize the diagnoses that the agents actually receive.

For the repetitions of \Cref{rq1}, we additionally report the resolved instances of each run and their mean over the runs of an agent and mode.
We compare every \FORECAST{} run of an agent with every baseline run of the same agent with the McNemar test, as in the main experiment.
These are nine comparisons per agent with \QWEN{} and two with \HAIKU{}.
With these $55$ comparisons, we again apply the Bonferroni correction.
Since parts of the repair agents are non-deterministic, two runs of the same mode may also disagree on some instances.
One run resolves an instance that the other does not, and vice versa.
The number of such instances shows how much the outcome varies from run to run by chance alone.
We judge the difference between a variant and the \FORECAST{} mode against this variation.
For \Cref{rq2}, we compare each variant of an agent with each of the agent's three \FORECAST{} runs on the paired per-instance outcomes, again with the exact McNemar test, and report its change against their mean.
With $23$ variants and agents ($13$ with \MINISWEAGENT{} and $10$ with \REPAIRAGENT{}) and three runs each, we apply the Bonferroni correction over $69$ comparisons.
We consider a variant significantly different from the \FORECAST{} mode if it differs significantly from each of its three runs.
A variant that resolves significantly fewer instances withholds or replaces information the agents benefit from.

\subsection{Subjects}%
\label{sec:subjects}

\begin{table}[t]
    \caption{Subjects and their diagnoses.
    Per project of \SWELITE{}: instances, traced failing and passing test executions, median number of causes per chain, chains with a discriminability of 1.00, and median time to compute a diagnosis.
    The last four columns give the share of resolved repair attempts over all five agents, without (Base) and with (\FORECAST{}) the diagnosis.}%
    \label{tab:subjects}
    \Description{A table with one row per project of SWE-bench Lite and a total row. It gives the number of instances, the traced failing and passing tests, the median number of causes per chain, the chains with a discriminability of 1.00, the median time to compute a diagnosis, and the share of resolved repair attempts without and with the diagnosis for both models. With the diagnosis, every project resolves more attempts with both models, except flask with Qwen3.5, which stays at 20 percent.}
    \small
    \begin{tabular}{@{}lrrrrrrrrrr@{}}
    \toprule
    & & \multicolumn{2}{c}{\textbf{Traced tests}} & & & & \multicolumn{2}{c}{\textbf{\HAIKUSHORT{} (\%)}} & \multicolumn{2}{c@{}}{\textbf{\QWEN{} (\%)}} \\
    \cmidrule(lr){3-4} \cmidrule(lr){8-9} \cmidrule(l){10-11}
    \textbf{Project} & \textbf{Inst.} & \textbf{Fail} & \textbf{Pass} & \textbf{Causes} & \textbf{Disc.\,1.00} & \textbf{Time (min)} & \textbf{Base} & \textbf{\FORECAST{}} & \textbf{Base} & \textbf{\FORECAST{}} \\
    \midrule
    astropy      &   6 &     8 &    217 & 10 &   6 &  33 & 43.3 & 73.3 & 33.3 & 50.0 \\
    \rowcolor{tablerow}django       & 114 &   716 &  5,841 & 10 & 102 &  45 & 58.6 & 79.6 & 36.0 & 62.8 \\
    flask        &   3 &     5 &    122 &  2 &   2 &   4 & 13.3 & 40.0 & 20.0 & 20.0 \\
    \rowcolor{tablerow}matplotlib   &  23 &    29 &  2,010 &  6 &  22 & 108 & 45.2 & 75.7 & 33.0 & 64.3 \\
    pylint       &   6 &     8 &    208 &  4 &   6 &   6 & 33.3 & 56.7 & 23.3 & 40.0 \\
    \rowcolor{tablerow}pytest       &  17 &    35 &  1,300 & 10 &  17 &  53 & 38.8 & 71.8 & 27.1 & 54.1 \\
    requests     &   6 &    42 &    556 & 10 &   6 &   9 & 36.7 & 43.3 & 23.3 & 40.0 \\
    \rowcolor{tablerow}scikit-learn &  23 &    55 &  1,199 &  7 &  20 &   9 & 58.3 & 74.8 & 47.0 & 67.0 \\
    seaborn      &   4 &     4 &    256 &  3 &   4 &   6 & 60.0 & 70.0 & 45.0 & 75.0 \\
    \rowcolor{tablerow}sphinx       &  16 &    23 &    388 &  9 &  11 &   5 & 37.5 & 62.5 & 27.5 & 42.5 \\
    sympy        &  77 &   100 &  4,231 & 10 &  73 &  76 & 39.7 & 62.1 & 22.1 & 47.5 \\
    \rowcolor{tablerow}xarray       &   5 &    15 &    392 & 10 &   5 &  12 & 40.0 & 56.0 & 20.0 & 52.0 \\
    \midrule
    \textbf{Total} & \textbf{\nInstances} & \textbf{\nFailingTraced} & \textbf{\nPassingTraced} & \textbf{10} & \textbf{\nPerfectChains} & \textbf{\diagMedianMinutes} & \textbf{48.5} & \textbf{70.9} & \textbf{31.2} & \textbf{56.1} \\
    \bottomrule
    \end{tabular}
\end{table}

\Cref{tab:subjects} summarizes our subjects, the $\nInstances$ instances of \SWELITE{} from $\nProjects$ projects, and the cause-chain diagnoses \FORECAST{} derives for them.
Django and SymPy alone contribute $191$ instances, whereas six projects contribute six or fewer.
For $299$ instances, the diagnosis is a non-empty cause chain.
For the remaining one (\texttt{pallets\_\_flask-4992}), no traced feature separates the failing from the passing executions.
The agents receive the corresponding note instead (\Cref{sec:enhancing-repair-agents}).
Most cause chains reach the maximal length of ten causes (a median of $10$, with quartiles of $5$ and $10$).
In $\nPerfectChains$ instances ($91\%$), the strongest cause separates the failing from the passing executions perfectly, with a discriminability of $1.00$.
With a median of one failing test per instance, such a perfect separation means that the cause holds in every failing and in no passing observation at its node.
It does not mean that the cause would separate further runs as well.
Only four cause chains are weak in the sense of \Cref{sec:enhancing-repair-agents}, reaching at most $0.1$, and they are presented to the agents with the corresponding caveat.
Nine more reach between $0.1$ and the minimal discriminability of $0.6$.
The simplification keeps their causes (\Cref{sub:simplifying}), and they are presented without a caveat.
The cause chains contain $2{,}367$ causes in total.
Of these, $1{,}445$ ($61\%$) are single features, $238$ ($10\%$) are conjunctions or disjunctions of features, and $684$ ($29\%$) are divergences, calls that only failing runs make.
Half of the cause chains ($147$) contain at least one divergence.

A cause-chain diagnosis describes how the failing execution differs from the passing ones.
It does not predict where the fix goes, and the two need not coincide.
To characterize how close the diagnoses come to the fix nonetheless, we compare them with the developer patches of \SWELITE{}, which \FORECAST{} itself never sees.
In $\nFixedFileChains$ instances ($55\%$), some cause of the cause chain is located in a file that the developer patch changes.
In $37$ instances ($12\%$), a cause names a line within one of the patch's hunks.
Naming code that the fix changes, however, is not the same as stating the fix.
We therefore inspected every cause chain with a cause that states a condition on a line the developer patch edits or a value that only the lines the patch adds contain.
Four cause chains contain such a cause, and in none of them does the cause state the fix, that is, the new behavior that the patch introduces.

\subsection{\Cref{rq1}: Effectiveness}%
\label{sec:rq1}

\begin{table}[t]
    \caption{Resolved instances of \SWELITE{} ($n = \nInstances$) per agent and model, without (Baseline) and with (\FORECAST{}) the diagnosis.
    $\Delta$: change in percentage points relative to the baseline, with the $95\%$ bootstrap confidence interval of the former.
    \emph{Gained} and \emph{Lost}: instances only the \FORECAST{} mode or only the baseline resolves.
    Underlined changes are significant ($p < 0.05$).}%
    \label{tab:rq1}
    \Description{A table with one row per agent and model and a total row per model. It gives the resolved instances without and with the diagnosis, the change in percentage points and relative to the baseline, the confidence interval of the change, and the instances gained and lost. Every change is positive and significant. Over all agents, the share of resolved instances rises from 48.5 to 70.9 percent with Claude Haiku 4.5 and from 31.2 to 56.1 percent with Qwen3.5.}
    \small
    \begin{tabular}{@{}llrrrrcrr@{}}
    \toprule
    \textbf{Model} & \textbf{Agent} & \textbf{Baseline} & \textbf{\FORECAST{}} & \multicolumn{2}{c}{\textbf{$\Delta$}} & \textbf{95\% CI} & \textbf{Gained} & \textbf{Lost} \\
    \midrule
    & \MINISWEAGENT{}      & 168 (56.0\%) & 201 (67.0\%) & \underline{$+11.0$} & \underline{$+20\%$} & [$+6.7$, $+15.7$] &  43 & 10 \\
    \rowcolor{tablerow}\cellcolor{white}& \SWEAGENT{}          & 160 (53.3\%) & 214 (71.3\%) & \underline{$+18.0$} & \underline{$+34\%$} & [$+13.0$, $+23.0$] &  61 &  7 \\
    & \AUTOCODEROVER{}     & 179 (59.7\%) & 210 (70.0\%) & \underline{$+10.3$} & \underline{$+17\%$} & [$+6.0$, $+15.0$] &  40 &  9 \\
    \rowcolor{tablerow}\cellcolor{white}& \REPAIRAGENT{}       & 133 (44.3\%) & 268 (89.3\%) & \underline{$+\raGainHaiku$} & \underline{$+102\%$} & [$+39.3$, $+50.7$] & 136 &  1 \\
    & \AGENTLESS{}         &  87 (29.0\%) & 170 (56.7\%) & \underline{$+27.7$} & \underline{$+95\%$} & [$+22.0$, $+33.7$] &  91 &  8 \\
    \cmidrule(l){2-9}
    \rowcolor{tablerow}\cellcolor{white}\multirow{-6}{*}{\HAIKUSHORT{}} & \textbf{All agents}  & \textbf{727 (48.5\%)} & \textbf{1,063 (70.9\%)} & \underline{$\mathbf{+\haikuGain}$} & \underline{$\mathbf{+\haikuGainRel}$} & [$+19.6$, $+25.2$] & \textbf{\haikuGained} & \textbf{\haikuLost} \\
    \midrule
    & \MINISWEAGENT{}      & 121 (40.3\%) & 196 (65.3\%) & \underline{$+25.0$} & \underline{$+62\%$} & [$+20.0$, $+30.0$] &  78 &  3 \\
    \rowcolor{tablerow}\cellcolor{white}& \SWEAGENT{}          &  69 (23.0\%) & 163 (54.3\%) & \underline{$+31.3$} & \underline{$+136\%$} & [$+26.0$, $+36.7$] &  95 &  1 \\
    & \AUTOCODEROVER{}     & 133 (44.3\%) & 193 (64.3\%) & \underline{$+20.0$} & \underline{$+45\%$} & [$+15.3$, $+24.7$] &  62 &  2 \\
    \rowcolor{tablerow}\cellcolor{white}& \REPAIRAGENT{}       &  50 (16.7\%) & 170 (56.7\%) & \underline{$+\raGainQwen$} & \underline{$+240\%$} & [$+34.3$, $+45.7$] & 121 &  1 \\
    & \AGENTLESS{}         &  95 (31.7\%) & 120 (40.0\%) & \underline{$+8.3$} & \underline{$+26\%$} & [$+4.7$, $+12.3$] &  31 &  6 \\
    \cmidrule(l){2-9}
    \rowcolor{tablerow}\cellcolor{white}\multirow{-6}{*}{\QWEN{} 27B} & \textbf{All agents}  & \textbf{468 (31.2\%)} & \textbf{842 (56.1\%)} & \underline{$\mathbf{+\qwenGain}$} & \underline{$\mathbf{+\qwenGainRel}$} & [$+22.3$, $+27.5$] & \textbf{\qwenGained} & \textbf{\qwenLost} \\
    \bottomrule
    \end{tabular}
\end{table}

\Cref{tab:rq1} contrasts every agent with and without the cause-chain diagnosis.
With either model, every one of the five agents resolves more instances when it receives the diagnosis.
All ten differences are significant ($p < 0.05$ after Bonferroni correction), as are the totals over all agents.
The gains far outnumber the losses.
Over all agents, the diagnosis lets an agent resolve $\haikuGained$ (\HAIKU{}) and $\qwenGained$ (\QWEN{}) instances that it does not resolve without it.
It costs an agent only $\haikuLost$ and $\qwenLost$ instances that it resolves without it.
For every instance an agent loses with the diagnosis, it thus gains about $\haikuGainsPerLoss$ (\HAIKU{}) or $\qwenGainsPerLoss$ (\QWEN{}).

\paragraph{Lost Instances}
To see why an agent fails to repair an instance with the cause-chain diagnosis, we inspected all $48$ such runs, $35$ with \HAIKU{} and $13$ with \QWEN{}.
In $15$ of them, the agent fixes only part of the problem, for instance, one of two affected methods, and in $13$, it produces no patch.
In $8$, it changes the right place in the wrong way, and in $7$, it changes another place than the developer patch.
Three runs fail only on a flaky test that the patch does not touch.
In $28$ of the $48$ runs, the cause chain names a file the developer patch changes.
This is about as often as for all instances ($55\%$), so the losses are not concentrated on cause chains that point elsewhere.

\paragraph{Test Access}
The cause-chain diagnosis is derived from the developer's failing and passing tests, which \SWEAGENT{}, \MINISWEAGENT{}, and \AGENTLESS{} never see.
Their gains could therefore merely reflect information from tests rather than the value of the cause chain.
We include \REPAIRAGENT{} and \AGENTLESS{} to control for this (\Cref{sec:setup}).
In both modes, \REPAIRAGENT{} runs exactly the developer's tests as often as its budget allows.
It is also told which methods spectrum-based fault localization over the same executions ranks highest.
Still, \REPAIRAGENT{} gains most of all agents, $\raGainHaiku$ percentage points with \HAIKU{} and $\raGainQwen$ with \QWEN{}, and loses a single instance with each model.
\AUTOCODEROVER{} also validates its patches against these tests and receives the same fault localization, and it gains $10.3$ and $20.0$ percentage points.
For both agents, these tests also decide whether an instance is resolved.
Their absolute rates are therefore not comparable to those of the other agents, although this holds for both of their modes alike.
Without the diagnosis, \REPAIRAGENT{} finds a fix that passes the tests within its budget of $40$ commands for $133$ and $50$ instances.
With the diagnosis, it finds one for $270$ and $172$ instances, of which the official evaluation confirms $268$ and $170$.
Its gain could thus partly reflect that the diagnosis lets it reach a passing fix within its budget.
Without the diagnosis, however, it finds most of its fixes early, after a median of $12$ (\HAIKU{}) and $9$ (\QWEN{}) of its $40$ commands.
Only $15$ of $133$ and $6$ of $50$ fixes come after the $30$th command, so a larger budget would likely add few fixes.
Without \REPAIRAGENT{}, the share of resolved instances over the other four agents rises from $49.5\%$ to $66.3\%$ with \HAIKU{} and from $34.8\%$ to $56.0\%$ with \QWEN{}.
Without both \REPAIRAGENT{} and \AUTOCODEROVER{}, it rises from $46.1\%$ to $65.0\%$ and from $31.7\%$ to $53.2\%$, each significantly (sign test, $p < 0.05$).
\AGENTLESS{} has no developer test, but in both modes, it generates a test that reproduces the issue and runs it on every candidate patch.
It thus controls for having a reproduction test, although the diagnosis may also improve the tests it generates.
\AGENTLESS{} still gains $27.7$ and $8.3$ percentage points.
The most direct control is the test-output variant, which gives \MINISWEAGENT{} the failing tests and their output instead of the cause chain.
It recovers only $40\%$ of the gain (\Cref{sec:rq2}).
The benefit thus does not stem from access to tests alone.

\paragraph{Localization}
The cause-chain diagnosis also helps where it does not point to the code the developer fixed.
Over all agents, the instances whose cause chain names a file the developer patched ($\nFixedFileChains$, \Cref{sec:subjects}) gain $23.7$ (\HAIKU{}) and $26.6$ (\QWEN{}) percentage points.
The remaining $134$ instances gain $20.7$ and $22.8$ percentage points.
A cause chain thus helps even when none of its causes lie in the file that needs to change.
We attribute this to what the chain conveys besides locations.
It shows which code the failing test exercises, where its execution departs from the passing ones, and which values and branches differ.
That narrows down where to look and what to look for, even when the fault-triggering code itself lies elsewhere along the way.
Consistently, the agents need fewer tokens per resolved instance with the diagnosis in nine of the ten configurations (\Cref{sec:rq4}).

\paragraph{\AGENTLESS{}}
The gain of \AGENTLESS{} depends strongly on the model.
With \HAIKU{}, it gains $27.7$ percentage points, the second largest gain with this model, but with \QWEN{}, it gains only $8.3$, the smallest of all.
Some instances end with an internal error in at least one of its configurations.
On the others ($286$ with \HAIKU{} and $292$ with \QWEN{}), it resolves $85$ without and $170$ with the cause-chain diagnosis, and $95$ and $120$, and both differences are significant ($p < 0.05$).
Its fixed pipeline localizes from the repository structure and samples its patches in a single pass.
We suspect that this leaves the weaker model little room to follow the diagnosis once a step has gone astray.

\paragraph{Success Rate}
Over all five agents, the cause-chain diagnosis raises the share of resolved repair attempts from $\haikuBase$ to $\haikuForecast$ with \HAIKU{} ($+\haikuGain$ percentage points, $+\haikuGainRel$ relative).
With \QWEN{}, it raises the share from $\qwenBase$ to $\qwenForecast$ ($+\qwenGain$ percentage points, $+\qwenGainRel$ relative, \Cref{tab:rq1}).
Per agent, the absolute gains range from $\minAgentGain$ to $\maxAgentGain$ percentage points, and the relative ones from $17\%$ to $240\%$.

\paragraph{Weaker Model, Larger Gains}
The cause-chain diagnosis helps the weaker model more.
\QWEN{} gains more than \HAIKU{} with three of the five agents, and relatively more with four.
The diagnosis more than compensates for the weaker model.
Four of the five agents resolve clearly more instances with \QWEN{} and the diagnosis than with \HAIKU{} alone.
These are \MINISWEAGENT{} ($65.3\%$ against $56.0\%$), \AUTOCODEROVER{} ($64.3\%$ against $59.7\%$), \REPAIRAGENT{} ($56.7\%$ against $44.3\%$), and \AGENTLESS{} ($40.0\%$ against $29.0\%$).
\SWEAGENT{} resolves about as many ($54.3\%$ against $53.3\%$), a difference within the variation between repeated runs.
Over all agents, \QWEN{} with the diagnosis ($\qwenForecast$) surpasses \HAIKU{} without it ($\haikuBase$), and it does so without \REPAIRAGENT{} as well ($56.0\%$ against $49.5\%$).
A locally served open-weight model with \FORECAST{} thus may repair more than a commercial model without it.

\paragraph{Agents}
The cause-chain diagnosis also reorders the agents.
Without the diagnosis, \REPAIRAGENT{} resolves fewer instances than the other autonomous agents.
With it and \HAIKU{}, \REPAIRAGENT{} becomes the most effective of all agents, resolving $89.3\%$ of the instances.
Its loop alternates between hypotheses about the bug and test runs that check them.
This loop may be particularly suited to a diagnosis that supplies such hypotheses.
\SWEAGENT{} with \QWEN{} more than doubles its rate, from $23.0\%$ to $54.3\%$.
That agent, however, rarely ends a run by itself with \QWEN{}.
In \FORECAST{} mode, $254$ of its $\nInstances$ runs (against $223$ in the baseline) end at the budget of about three million tokens.
\SWEAGENT{} then submits the changes made so far.
With the diagnosis, $129$ of these submissions resolve the instance, against $10$ in the baseline.
In these runs, the agent made a resolving change before the budget ran out but did not end the run.

\paragraph{Instances}
With \HAIKU{}, at least one of the five agents resolves $226$ instances ($75.3\%$) without and $283$ ($94.3\%$) with the cause-chain diagnosis.
With \QWEN{}, at least one agent resolves $180$ ($60.0\%$) and $248$ ($82.7\%$).
With the diagnosis, the agents resolve $57$ (\HAIKU{}) and $70$ (\QWEN{}) instances that no agent resolves without it.
Conversely, there are no such instances with \HAIKU{} and two with \QWEN{}.
The gains are not confined to some projects either.
Across all agents, the diagnosis raises the resolution rate in each of the $\nProjects$ projects with \HAIKU{}, and in $11$ with \QWEN{}, where the three Flask instances stay unchanged (\Cref{tab:subjects}).

\begin{table}[t]
    \caption{Resolved instances in every run of an agent and mode, the main experiment as run~1, and their mean.
    The smallest and largest gains, in instances, over every pair of a \FORECAST{} and a baseline run of the agent are underlined, where the gain is significant for every pair ($p < 0.05$).}%
    \label{tab:repeated}
    \Description{A table with the resolved instances of every run of each agent and model, with up to three baseline and three FORECAST runs and their means, and the smallest and largest gain over all pairs of a FORECAST and a baseline run. All gains are significant and range from 23 to 137 instances.}
    \small
    \begin{tabular}{@{}llrrrrrrrrrr@{}}
    \toprule
    & & \multicolumn{4}{c}{\textbf{Baseline}} & \multicolumn{4}{c}{\textbf{\FORECAST{}}} & \multicolumn{2}{c}{\textbf{Gain}} \\
    \cmidrule(lr){3-6}\cmidrule(lr){7-10}\cmidrule(l){11-12}
    \textbf{Model} & \textbf{Agent} & \textbf{1} & \textbf{2} & \textbf{3} & \textbf{Mean} & \textbf{1} & \textbf{2} & \textbf{3} & \textbf{Mean} & \textbf{Min.} & \textbf{Max.} \\
    \midrule
    & \MINISWEAGENT{} & 168 &  &  & 168.0 & 201 & 202 &  & 201.5 & \underline{$+33$} & \underline{$+34$} \\
    \rowcolor{tablerow}\cellcolor{white}& \SWEAGENT{} & 160 &  &  & 160.0 & 214 & 217 &  & 215.5 & \underline{$+54$} & \underline{$+57$} \\
    & \AUTOCODEROVER{} & 179 &  &  & 179.0 & 210 & 212 &  & 211.0 & \underline{$+31$} & \underline{$+33$} \\
    \rowcolor{tablerow}\cellcolor{white}& \REPAIRAGENT{} & 133 &  &  & 133.0 & 268 & 270 &  & 269.0 & \underline{$+135$} & \underline{$+137$} \\
    \multirow{-5}{*}{\HAIKUSHORT{}} & \AGENTLESS{} & 87 &  &  & 87.0 & 170 & 171 &  & 170.5 & \underline{$+83$} & \underline{$+84$} \\
    \midrule
    & \MINISWEAGENT{} & 121 & 119 & 114 & 118.0 & 196 & 198 & 201 & 198.3 & \underline{$+75$} & \underline{$+87$} \\
    \rowcolor{tablerow}\cellcolor{white}& \SWEAGENT{} & 69 & 72 & 73 & 71.3 & 163 & 167 & 168 & 166.0 & \underline{$+90$} & \underline{$+99$} \\
    & \AUTOCODEROVER{} & 133 & 130 & 133 & 132.0 & 193 & 196 & 192 & 193.7 & \underline{$+59$} & \underline{$+66$} \\
    \rowcolor{tablerow}\cellcolor{white}& \REPAIRAGENT{} & 50 & 49 & 56 & 51.7 & 170 & 175 & 174 & 173.0 & \underline{$+114$} & \underline{$+126$} \\
    \multirow{-5}{*}{\QWEN{} 27B} & \AGENTLESS{} & 95 & 90 & 97 & 94.0 & 120 & 122 & 122 & 121.3 & \underline{$+23$} & \underline{$+32$} \\
    \bottomrule
    \end{tabular}
\end{table}

\paragraph{Repeated Runs}
In the main experiment, each configuration ran once per instance.
We therefore examine how much a second and a third run differ from the first, and whether the gain of the cause-chain diagnosis persists in each of them (\Cref{sec:setup}).
\Cref{tab:repeated} lists the resolved instances of every run.
Runs of the same agent and mode differ by at most $7$ resolved instances.
Over all agents and averaged over its three runs, \QWEN{} resolves $31.1\%$ of the instances without the diagnosis and $56.8\%$ with it, against $31.2\%$ and $56.1\%$ in the main experiment.
With \HAIKU{}, the repeated \FORECAST{} mode resolves $71.5\%$ of the instances over all agents, against $70.9\%$ in the main experiment.
In all $55$ comparisons, the \FORECAST{} run resolves significantly more instances than the baseline run of the same agent ($p < 0.05$ after Bonferroni correction).
The gains range from $23$ instances (\AGENTLESS{} with \QWEN{}) to $137$ (\REPAIRAGENT{} with \HAIKU{}).
Two runs of the same mode nonetheless disagree on individual instances.
With \QWEN{}, two such runs of an agent disagree on $16$ to $73$ instances, and with \HAIKU{} on $12$ to $29$.
In the \FORECAST{} mode with \QWEN{}, for instance, they disagree on up to $73$ instances with \MINISWEAGENT{} and $52$ with \REPAIRAGENT{}.
With \HAIKU{}, the gains rest on a single baseline run.

\begin{finding}[RQ1: Every agent resolves significantly more issues with the cause-chain diagnosis, in every run]
With either model, all five agents repair significantly more instances of \SWELITE{} with the cause-chain diagnosis than without it.
The share of resolved instances rises from $\haikuBase$ to $\haikuForecast$ with \HAIKU{} and from $\qwenBase$ to $\qwenForecast$ with \QWEN{}, and without \REPAIRAGENT{} from $49.5\%$ to $66.3\%$ and from $34.8\%$ to $56.0\%$.
For each instance the agents lose with the diagnosis, they gain about $\haikuGainsPerLoss$ (\HAIKU{}) and $\qwenGainsPerLoss$ (\QWEN{}).
The gain is significant between every \FORECAST{} run and every baseline run of each agent in our repetitions.
The benefit persists for agents that already run the very tests the diagnosis is derived from, and for cause chains that do not point at the fixed file.
With the diagnosis, the open-weight model repairs more than the commercial model without it.
\end{finding}

\subsection{\Cref{rq2}: Diagnosis Components}%
\label{sec:rq2}

With the variants of \Cref{tab:variants}, we examine which parts of the cause-chain diagnosis the agents benefit from.
\Cref{tab:rq5} compares each variant with the three runs of the \FORECAST{} mode of the agent (\Cref{sec:rq1}).

\paragraph{Content}
\emph{Foreign}, \emph{Unordered}, and \emph{Locations} resolve significantly fewer instances than each \FORECAST{} run with both agents.
With the cause chain of another instance of the same project, \MINISWEAGENT{} resolves $119$ instances and \REPAIRAGENT{} $54$.
This is about as many as the $118.0$ and $51.7$ of their baseline runs on average, and neither differs significantly from any of these runs ($p \geq 0.05$).
A completely wrong diagnosis thus causes no detectable harm.
The agents gain nothing from it, but they also do not resolve fewer instances than without any diagnosis.
The same causes in a random order and without \texttt{LEADS TO} keep $40\%$ of the gain with \MINISWEAGENT{} and $37\%$ with \REPAIRAGENT{}.
The locations of the causes alone keep $9\%$ and $12\%$.
The prompts with a foreign cause chain are as long as those with the instance's own cause chain (a median of $2{,}305$ against $2{,}289$ characters).
Those with the locations alone are shorter ($1{,}892$), so the length of the prompt alone does not explain these losses.

\paragraph{Information Available without \FORECAST{}}
The failing tests and their output let \MINISWEAGENT{} resolve $150$ instances, $40\%$ of the gain of the cause chain.
The ranking of spectrum-based fault localization lets it resolve $126$ ($10\%$).
Both resolve significantly fewer than each run with the cause chain.
The test output is the most direct control for the information the cause-chain diagnosis draws from the developer's tests.
The cause chain followed by the test output lets \MINISWEAGENT{} resolve $234$ instances.
This is $35.7$ more than the cause chain alone on average and significantly more than each \FORECAST{} run.

\begin{wraptable}{r}{0.55\textwidth}
    \caption{Resolved instances with each variant of the diagnosis (\QWEN{}, $n = \nInstances$), and the change against the mean of the three \FORECAST{} runs of the agent (\Cref{tab:repeated}).
    Baseline and \FORECAST{} give the mean over their three runs.
    Underlined changes are significant against each of the three \FORECAST{} runs ($p < 0.05$).
    Instances where a variant does not change keep the outcome of the \FORECAST{} mode of the main experiment.}%
    \label{tab:rq5}
    \Description{A table with the resolved instances of mini-SWE-agent and RepairAgent with Qwen3.5 for each variant of the diagnosis, and the change against the mean of the three FORECAST runs. Foreign, Unordered, Locations, Test output, and SBFL resolve significantly fewer instances, Chain and test output resolves 35.7 more, and the remaining variants change the result by at most 13 instances.}
    \centering
    \small
    \begin{tabular}{@{}lrrrr@{}}
    \toprule
    & \multicolumn{2}{c}{\textbf{\MINISWEAGENT{}}} & \multicolumn{2}{c}{\textbf{\REPAIRAGENT{}}} \\
    \cmidrule(lr){2-3}\cmidrule(l){4-5}
    \textbf{Variant} & \textbf{Resolved} & \textbf{$\Delta$} & \textbf{Resolved} & \textbf{$\Delta$} \\
    \midrule
    Baseline              & 118.0 & & 51.7 & \\
    \rowcolor{tablerow}\FORECAST{}           & 198.3 & & 173.0 & \\
    \midrule
    Foreign               & 119 & \underline{$-79.3$} & 54 & \underline{$-119.0$} \\
    \rowcolor{tablerow}Unordered             & 150 & \underline{$-48.3$} & 96 & \underline{$-77.0$} \\
    Locations             & 125 & \underline{$-73.3$} & 66 & \underline{$-107.0$} \\
    \rowcolor{tablerow}No divergences        & 192 & $-6.3$ & 160 & $-13.0$ \\
    Atoms                 & 186 & $-12.3$ & 169 & $-4.0$ \\
    \rowcolor{tablerow}Test output           & 150 & \underline{$-48.3$} &  &  \\
    SBFL                  & 126 & \underline{$-72.3$} &  &  \\
    \rowcolor{tablerow}Chain and test output & 234 & \underline{$+35.7$} &  &  \\
    Wording               & 197 & $-1.3$ & 169 & $-4.0$ \\
    \rowcolor{tablerow}$k = 5$               & 195 & $-3.3$ & 172 & $-1.0$ \\
    $k = 20$              & 198 & $-0.3$ & 168 & $-5.0$ \\
    \rowcolor{tablerow}$t = 0.4$             & 199 & $+0.7$ & 167 & $-6.0$ \\
    $t = 0.8$             & 194 & $-4.3$ & 173 & $\pm0.0$ \\
    \bottomrule
    \end{tabular}
\end{wraptable}
\paragraph{Kinds of Causes, Wording, and Parameters}
The remaining variants differ from the mean of the \FORECAST{} runs by at most $13$ instances, and none differs significantly from any \FORECAST{} run.
Without divergences, \MINISWEAGENT{} resolves $6.3$ and \REPAIRAGENT{} $13.0$ fewer instances on average, and without conjunctions and disjunctions $12.3$ and $4.0$ fewer.
On the $147$ instances whose cause chain contains a divergence, \MINISWEAGENT{} resolves $94$ without divergences against $98$ to $102$ in the \FORECAST{} runs, and \REPAIRAGENT{} $71$ against $81$ to $85$.
On the $125$ instances whose cause chain contains a conjunction or disjunction, the agents resolve $74$ against $81$ to $84$ and $67$ against $68$ to $69$ without them.
On these affected instances alone, no difference is significant either (paired McNemar tests against each \FORECAST{} run, Bonferroni over $12$ tests for two agents, two variants, and three runs).
Still, \REPAIRAGENT{} without divergences resolves $10$ to $14$ fewer, a drop that is not negligible.
The experiment thus cannot resolve effects of this size.
It detects no effect of divergences and combinations, which does not show that they have none.
The neutral wording changes the number of resolved instances by $1.3$ and $4.0$ on average, and the other parameters of the simplification by at most $6$.

\paragraph{Discussion}
The variants thus support three properties of the cause-chain diagnosis.
It is specific to the failure at hand (\emph{Foreign}).
It states the conditions, values, and data flows of its causes rather than only their locations (\emph{Locations}).
It presents its causes in the order of execution (\emph{Unordered}).
\emph{Unordered} also drops the numbering and \texttt{LEADS TO}.
\emph{Wording}, however, keeps the order, replaces \texttt{LEADS TO} by \texttt{THEN}, and performs the same, so the loss stems from the order, not from the connector.
The order of the observations from the start of the test to the failure thus matters empirically.
This supports presenting them as a chain that leads to the failure, although it does not establish that each of them causes the next.
The further capabilities of \FORECAST{}, namely divergences, combinations of features, and the parameters of the simplification, make a cause-chain diagnosis more complete and more expressive.
Our experiments do not show, however, that they contribute to repair success on their own.
An unrelated diagnosis causes no detectable harm, whereas the effect of a related but misleading one remains untested.

\begin{finding}[RQ2: The benefit stems from what the cause chain says about the failure at hand]
With the cause chain of another instance, the agents resolve about as many instances as without any diagnosis, so an unrelated diagnosis neither helps nor causes detectable harm.
With the causes unordered or reduced to their locations, the agents lose most of the gain.
With \MINISWEAGENT{}, neither the failing tests' output nor a ranking of suspicious methods matches the cause chain.
Together with the test output, however, the cause chain lets \MINISWEAGENT{} resolve $35.7$ more instances on average.
Omitting divergences or combinations, wording the cause chain neutrally, or varying its parameters has no detectable effect.
\end{finding}

\subsection{\Cref{rq3}: Fix Quality}%
\label{sec:rq3}

\begin{table}[t]
    \caption{Generated patches.
    \emph{Patches}: runs that end with a non-empty patch.
    \emph{Applied}: share of these that apply cleanly.
    \emph{New files}: patches that also add files the agent created for itself.
    \emph{Changed lines}: added plus removed lines in existing files, median (first--third quartile), over all patches and over the instances both modes resolve (\emph{Both}).}%
    \label{tab:rq3}
    \Description{A table with, per model, agent, and mode, the number of patches, the share of them that applies, the patches that add new files, and the median number of changed lines over all patches and over the instances that both modes resolve. Applicability and patch size hardly differ between the modes. The exception is SWE-agent with Qwen3.5, whose patches add new files in 206 of 282 runs with the diagnosis.}
    \small
    \begin{tabular}{@{}lllrrrrr@{}}
    \toprule
    & & & & & & \multicolumn{2}{c@{}}{\textbf{Changed lines}} \\
    \cmidrule(l){7-8}
    \textbf{Model} & \textbf{Agent} & \textbf{Mode} & \textbf{Patches} & \textbf{Applied} & \textbf{New files} & \textbf{All patches} & \textbf{Both} \\
    \midrule
    &   & Baseline      & 299 & 100\% &   0 & 7 (4--15)  & 6 \\
    & \multirow{-2}{*}{\MINISWEAGENT{}} & \FORECAST{}   & 293 & 100\% &   0 & 6 (4--12)  & 5 \\
    \rowcolor{tablerow}\cellcolor{white}&       & Baseline      & 294 & 100\% &  12 & 9 (4--21)  & 7 \\
    \rowcolor{tablerow}\cellcolor{white}& \multirow{-2}{*}{\SWEAGENT{}} & \FORECAST{}   & 296 & 100\% &   4 & 8 (4--15)  & 6 \\
    &  & Baseline      & 293 & 100\% &   0 & 6 (3--10)  & 5 \\
    & \multirow{-2}{*}{\AUTOCODEROVER{}} & \FORECAST{}   & 269 & 100\% &   0 & 5 (3--8)   & 4 \\
    \rowcolor{tablerow}\cellcolor{white}&    & Baseline      & 133 & 100\% &   0 & 4 (2--8)   & 4 \\
    \rowcolor{tablerow}\cellcolor{white}& \multirow{-2}{*}{\REPAIRAGENT{}} & \FORECAST{}   & 270 & 100\% &   0 & 6 (3--10)  & 4 \\
    &      & Baseline      & 191 & 100\% &   0 & 5 (2--8)   & 4 \\
    \multirow{-10}{*}{\HAIKUSHORT{}} & \multirow{-2}{*}{\AGENTLESS{}} & \FORECAST{}   & 280 & 100\% &   0 & 6 (3--10)  & 4 \\
    \midrule
    &   & Baseline      & 217 & 97.2\% &   0 & 6 (2--16)  & 6 \\
    & \multirow{-2}{*}{\MINISWEAGENT{}} & \FORECAST{}   & 284 & 100\%  &   3 & 7 (4--11)  & 6 \\
    \rowcolor{tablerow}\cellcolor{white}&       & Baseline      & 110 & 100\%  &  15 & 4 (2--7)   & 4 \\
    \rowcolor{tablerow}\cellcolor{white}& \multirow{-2}{*}{\SWEAGENT{}} & \FORECAST{}   & 282 & 100\%  & 206 & 8 (3--19)  & 6 \\
    &  & Baseline      & 293 & 99.7\% &   0 & 6 (3--10)  & 5 \\
    & \multirow{-2}{*}{\AUTOCODEROVER{}} & \FORECAST{}   & 284 & 100\%  &   0 & 5 (3--10)  & 4 \\
    \rowcolor{tablerow}\cellcolor{white}&    & Baseline      &  50 & 100\%  &   0 & 2 (2--5)   & 2 \\
    \rowcolor{tablerow}\cellcolor{white}& \multirow{-2}{*}{\REPAIRAGENT{}} & \FORECAST{}   & 172 & 100\%  &   0 & 3 (2--5)   & 3 \\
    &      & Baseline      & 286 & 100\%  &   0 & 4 (2--7)   & 4 \\
    \multirow{-10}{*}{\QWEN{} 27B} & \multirow{-2}{*}{\AGENTLESS{}} & \FORECAST{}   & 290 & 100\%  &   0 & 4 (2--8)   & 4 \\
    \bottomrule
    \end{tabular}
\end{table}

\Cref{tab:rq3} characterizes the patches the agents produce.

\paragraph{Applicability}
Applicability hardly differs between the modes.
Every patch of every agent applies cleanly, except for six patches of \MINISWEAGENT{} and one of \AUTOCODEROVER{} with \QWEN{}, all in the baseline.
All five agents derive their patches from edits they have already applied to a working copy, or check that they apply.
A patch that does not apply is therefore rare in either mode.
What differs is whether an agent produces a patch at all.
\REPAIRAGENT{} only reports a fix that passes its tests.
With the cause-chain diagnosis, it produces a patch twice (\HAIKU{}) and more than three times (\QWEN{}) as often.
\QWEN{} with \SWEAGENT{} produces $282$ instead of $110$ patches, and with \MINISWEAGENT{} $284$ instead of $217$.
\AUTOCODEROVER{} is the only agent that produces fewer patches with the diagnosis ($269$ against $293$ with \HAIKU{}, $284$ against $293$ with \QWEN{}), yet more of them are correct.
With \HAIKU{}, $78\%$ of its patches resolve the instance with the diagnosis, against $61\%$ without it, and with \QWEN{}, $68\%$ against $45\%$.

\paragraph{Minimality}
The cause-chain diagnosis does not make the changes to existing code larger.
On the instances that both modes resolve, the median patches of the two modes differ by at most two lines for every agent and model.
In eight of ten configurations, the median patch with the diagnosis is the same size or smaller.
Over all patches, the medians differ by at most four lines, and the median patch modifies a single existing file in every configuration.
In some configurations, such as \REPAIRAGENT{} with \HAIKU{}, the median over all patches grows slightly with the diagnosis.
These medians include the many instances that only the \FORECAST{} mode resolves, which need not be the easiest ones.

The one notable difference concerns files the agents create for themselves.
\SWEAGENT{} instructs its model to write a script that reproduces the issue, and before it submits, it asks the model to remove such scripts again.
With \QWEN{} in \FORECAST{} mode, $206$ of $282$ patches still carry such files, namely test and debugging scripts and, in some cases, a summary of the fix.
These runs ended at their budget, where \SWEAGENT{} submits the current changes without that review (\Cref{sec:rq1}).
The files touch no existing code and do not affect the verdict.
A developer would still have to remove them, however, so they are a genuine cost of this configuration.
Counting them, the median patch of this configuration grows from $8$ to $139$ changed lines.
For all other configurations, new files are rare in both modes.

\paragraph{Semantic Correctness}
A patch that resolves an instance passes the tests of \SWELITE{}, but it may still miss what the developers intended.
We therefore examined two configurations, \MINISWEAGENT{} with \QWEN{} and \REPAIRAGENT{} with \HAIKU{}, which resolve the most instances.
For each, we drew $50$ of the instances that the \FORECAST{} mode resolves, uniformly at random with a fixed seed, and compared each patch with the developer patch.
One author judged every patch, not blind to the mode but with the same criteria in both modes.
The leakage inspection of \Cref{sec:subjects} followed the same protocol.
With \MINISWEAGENT{}, $43$ of the $50$ patches are semantically correct; that is, they behave as the developer patch does.
The other $7$ are plausible, passing the tests but missing part of the intent.
With \REPAIRAGENT{}, $45$ are correct and $5$ plausible, and in neither sample is a patch unrelated to the issue.
The baseline also resolves $30$ and $26$ of the sampled instances.
On these, the \FORECAST{} mode yields $24$ correct patches with either agent, and the baseline $25$ and $22$.
Of the patches for instances that only the \FORECAST{} mode resolves, $19$ of $20$ and $21$ of $24$ are correct.
We also inspected the cause chains of the same samples by hand.
In $16$ and $12$ instances, the cause chain contains a cause on the way to the fault.
In another $9$ instances each, it contains a cause in a file the developer patched, but away from the changed code.
About half of the sampled cause chains thus contain no cause near the fault, and yet the agents gain throughout.
The cause chain therefore does not help by pointing at the cause.
It helps by conveying which code the failing test exercises, where its run departs from the passing ones, and which values and branches differ (\Cref{sec:rq1}).

\begin{finding}[RQ3: The cause-chain diagnosis does not enlarge changes to existing code]
With the cause-chain diagnosis, patches apply as reliably as without it, change as few lines of existing code, and modify a single file in the median.
At the same time, agents produce a patch more often, and their patches more often resolve the issue.
The exception is \SWEAGENT{} with \QWEN{}, whose patches more often carry scripts it created for itself when its runs end at the budget.
In two samples of $50$ resolved instances, $43$ and $45$ patches are semantically correct, and none is unrelated to the issue.
\end{finding}

\subsection{\Cref{rq4}: Efficiency}%
\label{sec:rq4}

\begin{wrapfigure}{r}{0.47\textwidth}
    \centering
    \includegraphics[width=0.47\textwidth]{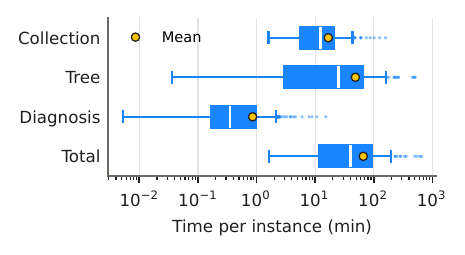}
    \caption{Wall-clock time to compute a diagnosis per instance, per step (collection includes instrumentation) and in total.
    Summed over all $\nInstances$ instances, the steps take $83.6$, $242.8$, $4.3$, and $330.7$ hours.}%
    \label{fig:rq4-diagnosis}
    \Description{Horizontal box plots on a logarithmic axis from 0.01 to 1,000 minutes, one each for the steps collection, tree, and diagnosis and for the total time per instance. Building the tree takes most of the time. Inferring the diagnosis takes a median of 21 seconds, and the total takes a median of 40 minutes.}
\end{wrapfigure}

\begin{figure}[t]
    \centering
    \includegraphics[width=\textwidth]{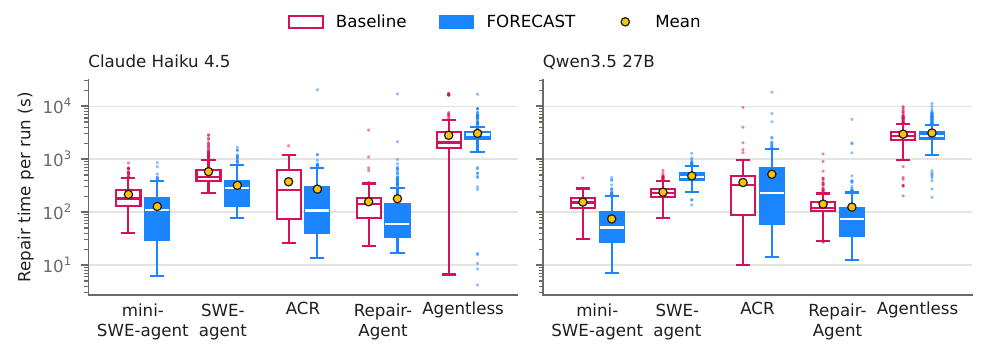}
    \caption{Repair time per run (logarithmic scale), per model and agent, without (Baseline) and with (\FORECAST{}) the diagnosis, excluding the time to compute the diagnosis.
    Boxes span the first to third quartile, with the median as a line and the mean as a yellow circle.
    Whiskers reach 1.5 times the interquartile range.
    ACR: \AUTOCODEROVER{}.}%
    \label{fig:rq4-time}
    \Description{Box plots of the repair time per run on a logarithmic axis in seconds, for each of the five agents with Claude Haiku 4.5 on the left and Qwen3.5 on the right, comparing the baseline with FORECAST. Agentless takes the longest, at about 50 minutes per run on average. With Claude Haiku 4.5, mini-SWE-agent and SWE-agent finish their runs faster with the diagnosis.}
\end{figure}

We weigh the cost of computing a cause-chain diagnosis (\Cref{fig:rq4-diagnosis}) against its effect on the tokens, cost, and time of the repair (\Cref{tab:rq4-repair,fig:rq4-time}).

\paragraph{Cost of the Diagnosis}
Computing a cause-chain diagnosis requires no LLM call.
It takes a median of $\diagMedianMinutes$ minutes of wall-clock time per instance (a mean of $66$ minutes), and less than an hour for $191$ of the $\nInstances$ instances.
The time varies strongly with the project.
The median is below ten minutes for six projects, but $76$ minutes for SymPy and $108$ minutes for Matplotlib.
For $21$ instances, it takes more than three hours, up to $10.9$ hours for the slowest one.
Tracing the tests accounts for a quarter of the total time.
Three quarters go to building the traced relevance tree, which merges all observations of all runs.
Inferring the cause chain from the finished tree takes a median of $21$ seconds.
Tree construction is thus the obvious target for optimization.
In our evaluation, it ran in a single process per instance on servers shared with other workloads, although passing runs can be folded in in parallel (\Cref{sec:implementation}).

For a single repair attempt, computing the diagnosis takes longer than the repair itself for all agents but \AGENTLESS{}, whose runs take about $50$ minutes on average (\Cref{fig:rq4-time}).
A cause-chain diagnosis, however, depends only on the failure, not on the agent or the model.
In our evaluation, each diagnosis served ten configurations.
In practice, it can serve any number of repair attempts, agents, or models, as well as the developer.
Moreover, computing it costs machine time, not developer time, and it can start as soon as a failing test exists, before any repair.

\begin{table}[t]
    \caption{LLM tokens and cost of the repair, per agent and model, without (Base) and with (\FORECAST{}) the diagnosis.
    Columns give the mean tokens per run, tokens per resolved instance, and, for \HAIKU{}, the cost at \ANTHROPIC{}'s list prices in total and per resolved instance.
    \QWEN{} runs locally and costs nothing per token.}%
    \label{tab:rq4-repair}
    \Description{A table with, per agent and model, the mean tokens per run, the tokens per resolved instance, and, for Claude Haiku 4.5, the cost in total and per resolved instance, each without and with the diagnosis. Over all agents, the tokens per resolved instance fall from 2.30 to 0.92 million with Claude Haiku 4.5 and from 3.05 to 1.51 million with Qwen3.5, and the cost per resolved instance falls from 0.85 to 0.49 dollars.}
    \small
    \begin{tabular}{@{}llrrrrrrrr@{}}
    \toprule
    & & \multicolumn{2}{c}{\textbf{Tokens / run}} & \multicolumn{2}{c}{\textbf{Tokens / resolved}} & \multicolumn{2}{c}{\textbf{Cost}} & \multicolumn{2}{c@{}}{\textbf{Cost / resolved}} \\
    \cmidrule(lr){3-4} \cmidrule(lr){5-6} \cmidrule(lr){7-8} \cmidrule(l){9-10}
    \textbf{Model} & \textbf{Agent} & \textbf{Base} & \textbf{\FORECAST{}} & \textbf{Base} & \textbf{\FORECAST{}} & \textbf{Base} & \textbf{\FORECAST{}} & \textbf{Base} & \textbf{\FORECAST{}} \\
    \midrule
    & \MINISWEAGENT{}        & 1.54M & 0.78M & 2.75M & 1.17M &  \$85.64 &  \$48.57 & \$0.51 & \$0.24 \\
    \rowcolor{tablerow}\cellcolor{white}& \SWEAGENT{}            & 3.16M & 1.74M & 5.93M & 2.44M & \$184.30 & \$100.54 & \$1.15 & \$0.47 \\
    & \AUTOCODEROVER{}       & 0.46M & 0.25M & 0.77M & 0.36M & \$183.75 & \$112.82 & \$1.03 & \$0.54 \\
    \rowcolor{tablerow}\cellcolor{white}& \REPAIRAGENT{}         & 0.21M & 0.15M & 0.48M & 0.17M &  \$74.05 &  \$65.01 & \$0.56 & \$0.24 \\
    & \AGENTLESS{}         & 0.19M & 0.35M & 0.64M & 0.61M &  \$93.67 & \$197.51 & \$1.08 & \$1.16 \\
    \cmidrule(l){2-10}
    \rowcolor{tablerow}\cellcolor{white}\multirow{-6}{*}{\HAIKUSHORT{}} & \textbf{All agents}    & \textbf{1.11M} & \textbf{0.66M} & \textbf{2.30M} & \textbf{0.92M} & \textbf{\$621.41} & \textbf{\$524.45} & \textbf{\$0.85} & \textbf{\$0.49} \\
    \midrule
    & \MINISWEAGENT{}        & 2.00M & 0.62M & 4.96M & 0.95M & -- & -- & -- & -- \\
    \rowcolor{tablerow}\cellcolor{white}& \SWEAGENT{}            & 2.19M & 2.84M & 9.52M & 5.24M & -- & -- & -- & -- \\
    & \AUTOCODEROVER{}       & 0.22M & 0.39M & 0.51M & 0.60M & -- & -- & -- & -- \\
    \rowcolor{tablerow}\cellcolor{white}& \REPAIRAGENT{}         & 0.17M & 0.18M & 1.04M & 0.32M & -- & -- & -- & -- \\
    & \AGENTLESS{}           & 0.17M & 0.21M & 0.54M & 0.52M & -- & -- & -- & -- \\
    \cmidrule(l){2-10}
    \rowcolor{tablerow}\cellcolor{white}\multirow{-6}{*}{\QWEN{} 27B} & \textbf{All agents}    & \textbf{0.95M} & \textbf{0.85M} & \textbf{3.05M} & \textbf{1.51M} & -- & -- & -- & -- \\
    \bottomrule
    \end{tabular}
\end{table}

\paragraph{Cost of the Repair}
On the LLM side, the cause-chain diagnosis makes most repairs cheaper (\Cref{tab:rq4-repair}).
With \HAIKU{}, \MINISWEAGENT{}, \SWEAGENT{}, and \AUTOCODEROVER{} need about half as many tokens per run, and \REPAIRAGENT{} needs more than a quarter less.
\MINISWEAGENT{} and \SWEAGENT{} also finish their runs in 40 to 45\% less time (\Cref{fig:rq4-time}).
Over all agents, the tokens per resolved instance fall by $\haikuTokenDrop$, from $2.3$ to $0.9$ million.
The LLM cost falls from \$$621$ to \$$524$ ($-16\%$).
Since the agents resolve more, the cost per resolved instance falls from \$$0.85$ to \$$0.49$ ($-\haikuCostDrop$).
The only exception, \AGENTLESS{}, nearly doubles its tokens per run.
With \QWEN{}, \MINISWEAGENT{} needs less than a third of the tokens per run and a fifth per resolved instance.
\SWEAGENT{}, in contrast, exhausts its budget in most runs in both modes, and in more runs with the diagnosis ($254$ against $223$, \Cref{sec:rq1}), so that it spends more.
\AUTOCODEROVER{}, too, spends more tokens with the diagnosis with \QWEN{} ($0.39$ against $0.22$ million per run on average), although fewer with \HAIKU{}.
Over all agents, \QWEN{} consumes $11\%$ fewer tokens with the diagnosis, and the tokens per resolved instance halve, from $3.1$ to $1.5$ million.

\paragraph{Amortization}
A cause-chain diagnosis is computed once per failure and serves every repair attempt for it.
In our main experiment alone, each diagnosis served the ten configurations of five agents and two models.
For these ten, the median of $\diagMedianMinutes$ minutes amounts to about $4$ minutes per configuration, against the full $\diagMedianMinutes$ minutes for a single repair attempt.

\begin{finding}[RQ4: The cause-chain diagnosis costs machine time but saves LLM tokens]
A cause-chain diagnosis costs a median of $\diagMedianMinutes$ minutes of machine time per failure, mostly for building the traced relevance tree, but no LLM tokens.
Any number of repair attempts can reuse it.
In return, the tokens per resolved instance fall in nine of ten configurations, over all agents, by $\haikuTokenDrop$ with \HAIKU{} and $\qwenTokenDrop$ with \QWEN{}.
With \HAIKU{}, the cost per resolved instance falls by $\haikuCostDrop$.
Whether these savings outweigh the machine time depends on how often a diagnosis is reused and on the price of computation.
\end{finding}

\section{Threats to Validity}%
\label{sec:threats-to-validity}

\subsection{Internal Validity}%
\label{sec:internal-validity}

\paragraph{Bounded Tracing.}
\FORECAST{} does not observe an execution in full.
It records a bounded prefix of large values, at most two iterations of any loop, and at most a fixed amount of trace per run (\Cref{sec:implementation}).
We chose the first two bounds so that the features we analyze are unaffected.
Value predicates are coarse, lengths are observed exactly, and loop features distinguish only never, once, and more than once.
The size budget, however, is a genuine loss.
A run that exceeds it is analyzed only up to that point, so a cause that manifests very late in a long execution can be missed.
This can only cost us causes, never invent them, so it works against \FORECAST{} rather than in its favor.

\paragraph{Warmed Environments.}
We perform each subject's one-time initialization before instrumenting it (\Cref{sec:implementation}).
The traced runs therefore no longer execute that initialization, and \FORECAST{} cannot observe it.
A fault that manifests only while a project sets itself up, for instance while it builds a cache, would be invisible to us, and the runs we compare would agree on it trivially.
We consider this an acceptable trade.
The alternative is to record the same initialization once per test, which in our subjects dominated both the size of the traces and the time to collect them.
Some subjects would then have been infeasible to evaluate at all.
As with the size budget, the effect is to withhold evidence from \FORECAST{}, not to supply it with any.

\paragraph{Test Selection.}
We do not trace every passing test, only those that execute code the failing tests also execute (\Cref{sec:implementation}).
\FORECAST{} compares executions only at functions that both a passing and a failing run reach.
Tests that share no code with the failing ones therefore cannot contribute to a cause-chain diagnosis, and dropping them cannot change the cause chains.
For instances with many passing tests, however, we also cap the number of tests whose coverage we measure ($200$) and that we trace ($120$, \Cref{sec:setup}).
The tests beyond these caps may share code with the failing ones, and their absence can change a cause chain.
Moreover, the selection relies on coverage measured once, and flaky or environment-dependent tests could be classified differently on another run.
The selection never consults the developer patch.

\paragraph{Data Contamination.}
The repositories and fixes of \SWEBENCH{} are public, and both models may have seen them during training.
This affects both modes of an agent alike and may raise their absolute resolution rates.
It does not explain, however, why the same agent with the same model resolves more with the cause-chain diagnosis than without it.
We cannot rule out that a cause chain pointing into familiar code helps a model recall a fix it has seen.
This would make the gains on a public benchmark larger than on unseen issues.

\subsection{External Validity}%
\label{sec:external-validity}

\paragraph{Subjects.}
We evaluate on the $\nInstances$ issues of \SWELITE{}, which come from $\nProjects$ Python projects, with five agents and two models.
Issues from other projects, issues that require changes across many files, other programming languages, and other models may benefit more or less from a cause-chain diagnosis.
The agents, however, differ widely in their designs, from fixed pipelines to autonomous loops, and the models differ in size and origin.
Every agent resolves significantly more issues with the diagnosis, with both models.
\SWELITE{} shares $93$ instances with the human-validated SWE-bench Verified.
On these, the share of resolved instances over all agents rises from $65.8\%$ to $86.7\%$ with \HAIKU{} and from $45.2\%$ to $73.3\%$ with \QWEN{}.
Since these are the same public instances, this speaks to the quality of the issue specifications, not to data contamination.
Both models run without thinking, and models that reason before they answer may benefit more or less from a cause-chain diagnosis.
The variants of the diagnosis ran only with \QWEN{} and two agents, and their results may differ for other agents and models.

\paragraph{Agent Fidelity.}
Two of our agents are not their authors' code.
\AUTOCODEROVER{} is completely re-implemented from its paper, and \REPAIRAGENT{} is ported from Java to Python.
Both follow their papers' designs and parameters.
Differences in prompts and tool behavior may still make them more or less effective than the originals, so their absolute resolution rates should not be compared with published ones.
This does not affect our comparison, which is between the two modes of the same implementation.
\SWEAGENT{}, \MINISWEAGENT{}, and \AGENTLESS{} run unmodified, apart from \AGENTLESS{}'s file retrieval, which we omit in both modes, and its larger answer length.

\paragraph{Availability of Failing Tests.}
\FORECAST{} derives its cause-chain diagnosis from failing and passing tests.
In our evaluation, the failing tests are the developer tests of \SWELITE{}.
\SWEAGENT{}, \MINISWEAGENT{}, and \AGENTLESS{} do not see them, so in the \FORECAST{} mode, the diagnosis is the only information they receive that stems from these tests.
\REPAIRAGENT{} and \AUTOCODEROVER{} run these tests in both modes.
The test-output variant shows the failing tests and their output instead of the cause chain.
Both show that access to these tests does not explain the benefit (\Cref{sec:rq1,sec:rq2}).
In practice, \FORECAST{} requires a test that reproduces the failure.
Where an issue comes without one, such a test has to be written first, by a developer or by a technique that generates it from the issue~\cite{kang2023libro, mundler2024swtbench}.
Generating such tests is a research problem of its own.
Evaluating diagnoses derived from generated tests is beyond the scope of this paper, which takes the reproducing test as given.

\subsection{Construct Validity}%
\label{sec:construct-validity}

\paragraph{Cost Accounting.}
We report costs only for \HAIKU{}, at \ANTHROPIC{}'s list prices.
\QWEN{}, which we serve locally, costs no fee per token, and we compare it by tokens.
The budgets of \SWEAGENT{} and \MINISWEAGENT{}, however, are defined in dollars.
For \QWEN{}, they therefore count its tokens at \HAIKU{}'s prices without prompt caching.
These budgets thus bind after fewer tokens with \QWEN{} than with \HAIKU{}.
This shapes the behavior of \SWEAGENT{} with \QWEN{} (\Cref{sec:rq1}), in both modes alike.
We measured the time to compute a cause-chain diagnosis on servers shared with other workloads, and it may be lower on dedicated hardware.

\paragraph{Resolution.}
We count an instance as resolved when the agent's patch makes the failing tests of \SWELITE{} pass and keeps the passing ones passing, as \SWEBENCH{} does.
A patch that passes these tests can still differ from what the developers intended, and a patch that fails them can still be a reasonable fix.
This affects both modes of an agent alike, and \Cref{sec:rq3} shows that the cause-chain diagnosis does not change the size or the applicability of the patches.
For two samples of $50$ resolved instances, we compared the patches with the developer patches by hand and found most of them semantically correct (\Cref{sec:rq3}), but we did not inspect every patch.
A single, unblinded author judged them so that the judgments may be biased, and the small samples allow counts, not precise rates.

\paragraph{Wording of the Diagnosis.}
The block presents the cause chain as a cause-effect chain and joins its causes by \texttt{LEADS TO} (\Cref{sec:enhancing-repair-agents}).
Only its causes are verified empirically, however, as observations that separate the runs.
That one cause brings about the next is not verified.
An agent may therefore trust a cause more than its evidence warrants.
Our main results measure the effect of the cause-chain diagnosis as worded.
The wording variant of \Cref{sec:rq2} presents the same chain in neutral terms, as observations joined by \texttt{THEN}.
It changes the number of resolved instances by at most $4$ on average, not significantly, whereas presenting the causes out of order loses most of the gain.
We therefore keep the term cause chain for a representation whose order matters, not as a claim that each of its causes brings about the next.
The agents also see the discriminability of every cause, which may likewise invite more trust than it warrants.
It reaches $1.00$ in $91\%$ of the cause chains, often because a single failing test makes a perfect separation easy.

\subsection{Conclusion Validity}%
\label{sec:conclusion-validity}

\paragraph{Nondeterminism.}
LLM-based agents are nondeterministic.
Some of them sample at a positive temperature, and even greedy decoding can vary with batching on the serving side.
The main experiment runs every configuration once per instance so that an individual instance may be resolved or not by chance.
The McNemar test compares the paired outcomes of the two modes.
It treats each outcome as fixed, however, and thus does not quantify how much a repeated run would differ.
We therefore repeat both modes of every agent twice with \QWEN{}, and the \FORECAST{} mode once with \HAIKU{} (\Cref{sec:rq1}).
With \HAIKU{}, the gains still rest on a single baseline run.
Each of the ten comparisons, however, is significant on its own after Bonferroni correction ($p < 0.05$).
Since all agents work on the same instances, their outcomes for an instance are not independent.
We therefore test the totals over all agents with a sign test over the instances, and they are significant with both models.
Each variant of the diagnosis ran once, and we compare it with each of the three runs of the \FORECAST{} mode.
Its difference to the \FORECAST{} mode can thus be judged against the variation between these runs.

\section{Related Work}%
\label{sec:related-work}

\paragraph{Test-based Program Repair}
Automated program repair generates patches that make the failing tests of a program pass while keeping the passing ones passing~\cite{legoues2019automated, monperrus2018automatic}.
GenProg evolves such patches by genetic programming from code that already exists in the program~\cite{weimer2009genprog, legoues2012genprog}.
SPR explores a staged search space of transformations and synthesizes the conditions they need~\cite{long2015spr}.
Prophet ranks the candidate patches by a model of correct code that it learns from past fixes~\cite{long2016prophet}.
SemFix and Angelix instead derive the constraints a fix has to satisfy by symbolic execution and synthesize code that satisfies them~\cite{nguyen2013semfix, mechtaev2016angelix}.
Getafix learns fix patterns from the fixes developers made before~\cite{bader2019getafix}, and neural approaches such as CoCoNuT translate buggy code into fixed code~\cite{lutellier2020coconut, zhang2023learning}.
For Python, FixKit makes several of these techniques available under a common interface~\cite{smytzek2024fixkit}.
A patch that passes the tests is not necessarily correct, however.
Tests specify the intended behavior only partially, and many plausible patches overfit to them~\cite{qi2015kali, smith2015cure}.
These techniques all start from fault localization, and how well they repair depends on how well it localizes~\cite{qi2013aprevaluation}.
\FORECAST{} gives a repair technique more than a location, namely an account of how the failing run departs from the passing ones.

\paragraph{LLM-based Program Repair}
Large language models have shifted automated program repair from templates and search to the generation of patches~\cite{xia2022alpharepair, xia2023llmapr} and to conversational repair that feeds test results back to the model~\cite{xia2024chatrepair}.
On repository-level benchmarks such as \SWEBENCH{}~\cite{jimenez2024swebench, openai2024verified}, repair is mostly performed by \emph{agents}.
\SWEAGENT{}~\cite{yang2024sweagent} and \MINISWEAGENT{}~\cite{sweagent2025minisweagent} let the model act on the repository through an agent-computer interface or a plain shell, and OpenHands~\cite{wang2025openhands} provides a general platform for such agents.
\AUTOCODEROVER{}~\cite{zhang2024autocoderover} retrieves context through code search and spectrum-based fault localization, and SpecRover~\cite{ruan2025specrover} additionally infers the intended behavior to guide and review its patches.
\REPAIRAGENT{}~\cite{bouzenia2025repairagent} drives the repair through hypotheses and test runs.
\AGENTLESS{}~\cite{xia2025agentless} shows that a fixed pipeline of localization, repair, and validation can compete with autonomous agents.
Other work trains models for this task~\cite{wei2025swerl}, studies how agents proceed~\cite{bouzenia2025trajectories}, or examines whether the issues they resolve are really solved~\cite{wang2026solved}.
\FORECAST{} proposes no new agent and instead improves the input of existing ones, for all five agents we evaluate.

\paragraph{Fault Localization}
Spectrum-based fault localization ranks program elements by how strongly their execution correlates with failing tests~\cite{jones2002tarantula, abreu2007accuracy, abreu2009practical, wong2016survey}, using formulas such as Ochiai~\cite{abreu2006ochiai} and DStar~\cite{wong2012dstar}.
It builds on the observation that differences in program spectra correlate with the exposure of faults~\cite{harrold2000empirical}.
Mutation-based fault localization instead mutates the program and observes which mutants change the outcome of the failing tests~\cite{papadakis2015metallaxis, moon2014muse}.
Later work enriches the spectra with further kinds of program behavior~\cite{santelices2009defuse, smytzek2026ffl}, such as the sequences of calls an object receives~\cite{dallmeier2005ample} or the likely invariants that differ between passing and failing runs~\cite{le2016savant}.
Other work learns rankings from many signals~\cite{li2019deepfl, zou2021combinefl} or from detailed coverage graphs~\cite{lou2021grace}, or it refines the ranking with the outcome of repair attempts~\cite{lou2020aprsfl}.
LLMs also localize faults, with or without tests~\cite{kang2024autofl, yang2024large, qin2026cosfl, chang2025llmfl}.
Results on artificial faults do not carry over to real ones~\cite{pearson2017evaluating}.
For Python, spectrum-based techniques have been evaluated on the real faults of BugsInPy~\cite{widyasari2022realworld}.
All of these techniques produce a ranking of locations, which repair tools consume as candidate edit locations.
Developers, however, profit little from such rankings alone~\cite{parnin2011automated}, and practitioners expect more than current techniques deliver~\cite{kochhar2016practitioners}.
A cause-chain diagnosis of \FORECAST{} instead tells how the failing execution differs from the passing ones.
In our evaluation, it helps agents even when it names no file the developer changed (\Cref{sec:rq1}).
This includes \AUTOCODEROVER{} and \REPAIRAGENT{}, which receive a spectrum-based ranking both with and without the diagnosis.

\paragraph{Slicing and Propagation}
Program slicing reduces a program to the statements that can influence a value at a given point~\cite{weiser1984program}, and dynamic slicing does so for a single execution~\cite{korel1988dynamic}.
A slice of the failing output contains the fault but is often large, and its effectiveness has been compared empirically with that of spectrum-based techniques~\cite{soremekun2021slicing}.
Other work follows how an infected program state propagates along the control flow towards the failure~\cite{zhang2009capturing}.
A cause chain of \FORECAST{} also follows the failure along the execution, but keeps only the steps in which the failing run differs from the passing ones and states how it differs.

\paragraph{Explaining Failures}
Model-based diagnosis explains the misbehavior of a system by the components whose failure is consistent with the observations~\cite{reiter1987diagnosis}.
Statistical debugging identifies the predicates over branches and return values whose outcome predicts failure~\cite{liblit2005statisticaldebugging, liu2005sober} and connects them to faulty control-flow paths~\cite{jiang2007contextaware}.
Dynamic invariants describe the relations among values that hold in passing runs~\cite{ernst2007daikon}.
Delta debugging isolates failure-inducing inputs~\cite{zeller2002deltadebugging}.
Applied to program states, it isolates the cause-effect chain from the input to the failure~\cite{zeller2002causeeffect, cleve2005causetransitions}, but this requires experiments that transplant states between a passing and a failing run.
BugEx generates tests that pass and fail close to a given failure and reports the branches and states that correlate with the failure~\cite{rossler2012bugex}.
Hypothesizer helps developers to find and test hypotheses about a failure~\cite{alaboudi2023hypothesizer}.
Input-centric approaches learn the properties of inputs under which a program fails~\cite{kampmann2020alhazen, eberlein2023avicenna}, and they require a specification of the input structure.
\FORECAST{} builds on the execution features of our prior work~\cite{smytzek2026efdd, smytzek2026javaefdd}.
This work showed that features such as conditions, values, and def-use pairs localize faults better than coverage alone and can also serve as failure oracles~\cite{smytzek2025bashiri}.
In \EFDD{}, the diagnosis is a decision tree over the features of whole runs.
\FORECAST{} instead compares the runs at every call and organizes the features that separate them into a cause chain along the calls of the program, using only the existing tests.
Like context-aware statistical debugging~\cite{jiang2007contextaware}, it thus connects discriminative predicates along the execution.
It does so over a call tree merged from all runs, however, and with features beyond branches and return values.
\FORECAST{} also approximates the cause-effect chains from input to failure that we proposed earlier~\cite{smytzek2023causeeffect}, without any experiments on the program state.
However, it does not verify that its causes cause one another.
We argued that such failure explanations should become a first-class artifact of program repair~\cite{smytzek2026diagnosis}.
\FORECAST{} realizes this idea for repair agents and measures its effect.

\paragraph{Execution Information for LLMs}
LLMs reason about program execution unreliably~\cite{gu2024cruxeval, liu2024codemind}, and several approaches therefore supply execution information to them.
Self-debugging lets a model explain its code and react to test results~\cite{chen2024selfdebug}.
LDB segments a single execution into blocks and lets the model check the intermediate values~\cite{zhong2024ldb}.
NExT and TRACED train models on execution traces~\cite{ni2024next, ding2024traced}.
AutoSD lets a model state hypotheses about a bug and test them with a debugger~\cite{kang2025autosd}, as InspectCoder does interactively~\cite{wang2025inspectcoder}.
DAIRA gives \SWEAGENT{} a tracer for the reproduction scripts the agent writes and has the model turn their call traces into structured reports~\cite{liu2026daira}.
These approaches expose single executions to the model and leave it to the model to find what is wrong.
ContrastRepair pairs a failing test with a similar passing one in the prompt~\cite{kong2025contrastrepair}, but it contrasts the two tests, not their executions.
\FORECAST{} instead contrasts all passing and failing executions it traces and hands the model only the differences, ranked by how sharply they separate the two.
Its cause-chain diagnosis is computed without any LLM, so every cause is an observation of the program rather than a narrative the model has invented.

\paragraph{Reproducing Failures}
Where an issue comes without a failing test, LLM-based test generation can reproduce it from the issue description~\cite{kang2023libro, mundler2024swtbench, ahmed2026otter, khatib2026assertflip, zheng2026reproagent}.
The generated tests can then serve as the failing tests from which \FORECAST{} derives its cause-chain diagnosis.
The passing tests that \FORECAST{} contrasts with the failing ones can be generated as well, by search-based test generators such as EvoSuite~\cite{fraser2011evosuite} and, for Python, Pynguin~\cite{lukasczyk2022pynguin}.
An LLM can help such generators past coverage plateaus~\cite{lemieux2023codamosa}.
Whether generated tests serve a diagnosis as well as the developer's tests do remains open.
Fault localization and repair already become less effective with the tests users provide than with those of the developer's fix~\cite{just2018developertests}.

\section{Conclusion and Future Work}%
\label{sec:conclusion}

Repair agents start from an issue description that tells them what goes wrong, but not how.
From the executions of passing and failing tests, it infers a cause chain of execution features that separate the failing from the passing runs.
It appends this cause-chain diagnosis to the issue the agent receives, without modifying the agent.
On the $\nInstances$ issues of \SWELITE{}, the diagnosis lets each of five repair agents of different designs resolve significantly more issues, with both a commercial and an open-weight model.
Over all agents, the share of resolved issues rises from $\haikuBase$ to $\haikuForecast$ with \HAIKU{} and from $\qwenBase$ to $\qwenForecast$ with \QWEN{}.
The gains persist across repeated runs, for agents that already run the tests the diagnosis is derived from, and for diagnoses that do not point at the code the developer changed.
They vanish when the cause chain is replaced by that of another issue, which causes no detectable harm either.
The changes to existing code remain as small as before, and the LLM tokens spent per resolved issue fall by $\haikuTokenDrop$ with \HAIKU{} and $\qwenTokenDrop$ with \QWEN{}.
With the diagnosis, the open-weight model resolves more issues than the commercial model without it.

These results open several directions for future work:
\begin{description}
    \item[Issues without failing tests.] In practice, an issue often comes without a test that reproduces it.
    Combining \FORECAST{} with test generation from issue descriptions~\cite{kang2023libro, mundler2024swtbench} would make it applicable to such issues.
    It would also let us study how the quality of generated tests affects the cause-chain diagnosis.
    \item[Faster diagnoses.] Building the traced relevance tree takes three-quarters of the time to compute a cause-chain diagnosis, and it ran in a single process in our evaluation.
    Folding in passing runs in parallel and tracing fewer but more informative tests could reduce the median of $\diagMedianMinutes$ minutes substantially.
    \item[Other languages and benchmarks.] Execution features also relate to faults in Java~\cite{smytzek2026javaefdd}.
    Porting \FORECAST{} to further languages and evaluating it on further benchmarks would show how far its benefits generalize.
    \item[Diagnoses beyond the prompt.] A cause chain describes how the failing execution departs from the passing ones, and a correct fix is likely to remove some of these differences.
    Checking whether a patch does so could validate candidate patches beyond the tests.
    The same cause-chain diagnosis can also help developers debug by hand.
\end{description}

\section*{Data Availability}%
\label{sec:tool}

\FORECAST{} and all evaluation scripts are open source under the Apache~2.0 license.
The current version is available at
\url{https://github.com/smythi93/forecast},
and \FORECAST{} can be installed as a package using \texttt{pip install forecast-repair}.
For long-term availability, we have archived the evaluation data, including every diagnosis and run, on Zenodo~\cite{forecast-artifact}.

\begin{acks}
    This research was partially funded by the Deutsche Forschungsgemeinschaft (DFG, German Research Foundation) --- ZE 509/7--2 and GR 3634/4--2 Emperor (261444241) and by the European Union (ERC S3, 101093186 and ERC PROSA, 101247971). Views and opinions expressed are, however, those of the authors only and do not necessarily reflect those of the European Union or the European Research Council. Neither the European Union nor the granting authority can be held responsible for them.
\end{acks}

\bibliographystyle{ACM-Reference-Format}
\bibliography{references}

\end{document}